\documentclass[prd,
 reprint,
superscriptaddress,
preprintnumbers,
nofootinbib,
 amsmath,amssymb,
 aps,
]{revtex4-1}
\usepackage[ocgcolorlinks]{hyperref}
\usepackage{graphicx}

\usepackage[utf8]{inputenc}
\usepackage[dvipsnames]{xcolor}
\usepackage{subcaption}
\usepackage{physics}

\newcommand{\MeV}{\text{\,MeV}}

\newcommand{\be}{\begin{equation}}
\newcommand{\ee}{\end{equation}}
\newcommand{\ba}{\begin{eqnarray}}
\newcommand{\ea}{\end{eqnarray}}

\newcommand{\la}{\langle}
\newcommand{\ra}{\rangle}

\def\ltap{\raisebox{-.55ex}{\rlap{$\sim$}} \raisebox{.4ex}{$<$}}

\def\lsim{\mathrel{\ltap}}

\begin{document}

\preprint{INR-TH-2025-0XX}

\title{
Searches for new light particles at the Troitsk Meson Factory (TiMoFey)} 

\author{Sergey Demidov}
\email{demidov@ms2.inr.ac.ru}
\affiliation{Institute for Nuclear Research of the Russian Academy of Sciences, 117312 Moscow, Russia}
\affiliation{Moscow Institute of Physics and Technology, 141700 Dolgoprudny, Russia}
\author {Alexander Feschenko}
\affiliation{Institute for Nuclear Research of the Russian Academy of Sciences, 117312 Moscow, Russia}
\author{Dmitry Gorbunov}
\email{gorby@ms2.inr.ac.ru}
\affiliation{Institute for Nuclear Research of the Russian Academy of Sciences, 117312 Moscow, Russia}
\affiliation{Moscow Institute of Physics and Technology, 141700 Dolgoprudny, Russia}
\author{Alexander Izmaylov}
\affiliation{Institute for Nuclear Research of the Russian Academy of Sciences, 117312 Moscow, Russia}
\author{Dmitry Kalashnikov}
\email{kalashnikov.d@phystech.edu}
\affiliation{Institute for Nuclear Research of the Russian Academy of Sciences, 117312 Moscow, Russia}
\affiliation{Moscow Institute of Physics and Technology, 141700 Dolgoprudny, Russia}
\author{Leonid Kravchuk}
\affiliation{Institute for Nuclear Research of the Russian Academy of Sciences, 117312 Moscow, Russia}
\author{Ekaterina Kriukova}
\email{kryukova.ea15@physics.msu.ru}
\affiliation{Institute for Nuclear Research of the Russian Academy of Sciences, 117312 Moscow, Russia}
\affiliation{Lomonosov Moscow State University, 119991 Moscow, Russia}
\author{Yury Kudenko}
\affiliation{Institute for Nuclear Research of the Russian Academy of Sciences, 117312 Moscow, Russia}
\affiliation{Moscow Institute of Physics and Technology, 141700 Dolgoprudny, Russia}
\affiliation{National Research Nuclear University MEPhI, 115409 Moscow, Russia}
\author{Nikita Mashin}
\affiliation{Institute for Nuclear Research of the Russian Academy of Sciences, 117312 Moscow, Russia}
\author{Yury Senichev}
\affiliation{Institute for Nuclear Research of the Russian Academy of Sciences, 117312 Moscow, Russia}
\affiliation{Moscow Institute of Physics and Technology, 141700 Dolgoprudny, Russia}

\date{\today}

\begin{abstract}
The project of a new accelerator complex at the Institute for Nuclear Research of RAS in Troitsk has recently been included in the Russian National Program ``Fundamental Properties of Matter". It will sustain a proton beam with a current of 300\,(100)\,$\mu$A and a proton kinetic energy of $T_p=423\,(1300)$\,MeV at the first\,(second) stage of operation. The complex is multidisciplinary, and here we investigate its prospects in exploring new physics with light, feebly interacting particles. We find that TiMoFey can access new regions of parameter space of models with light axion-like particles and models with hidden photons, provided by a generic multipurpose detector installed downstream the proton beam dump. The signature to be exploited is  the decay of a new particle into a pair of known particles inside the detector. Likewise, TiMoFey can probe previously unreachable ranges of parameters of models with millicharged particles obtained in measurements with detectors  recognizing energy deposits associated with elastic scattering of new particles, passing through the detector volume. The latter detector may be useful for dark matter searches, as well as for studies of neutrino physics suggested at the facility in the Program framework.     
\end{abstract}

\maketitle

\section{Introduction}
\label{sec:Introduction}
Explanations of dark matter, neutrino oscillations, baryon asymmetry of the Universe (and some other phenomenological issues with no means to be addressed within the Standard Model of particle physics) naturally imply an existence of new particles. They may be light or heavy. 
New light particles may interact only feebly with the Standard Model (SM) particles, and hence are best to be searched for in the intensity frontier experiments \cite{Essig:2013lka,Beacham:2019nyx,Agrawal:2021dbo,Antel:2023hkf}. Namely, the new particles may be produced by energetic beam particles scattering on a target. As far as the production of new particles is concerned, at fixed energy of a beam particle, the electron beams are favored by kinematic considerations, while the proton beams are more universal because of a variety of production mechanisms involved. 

In this paper we discuss a proposal of new experimental facility in Troitsk, at INR RAS, recently submitted to the project of Russian National Target Program ``Fundamental Properties of Matter". The key ingredient is a high-intensity proton beam with average current of 300(100)\,$\mu$A and proton kinetic energies of $T_p=423(1300)$\,MeV  at the two subsequent stages of operation. A possible future upgrade of the accelerating system with a Superconducting Linac may increase the current further by a factor of 3-5. 

The protons will hit the beam dump which stops them as well as secondary hadrons and charged leptons. The stopped charged pions decay, yielding a close-to-monochromatic muon neutrino flux, sufficiently intensive for investigations of neutrino coherent scattering in the nearby detectors. This is the main task of the Troitsk Meson Factory (or Troitsk Multipurpose Facility), which we call TiMoFey in short. 

At the same time, the incident protons may produce light hypothetical particles, if any. Being feebly interacting and long-lived, they may escape the dump and pass through the detectors providing with energy deposits, $e^+e^-$-pairs and other bright signatures. Specific neutrino detectors and general purpose detectors of photons, electrons and positrons will allow one to hunt for such signatures. The purpose of this paper is to estimate the TiMoFey sensitivity to a selected set of new physics models, predicting the existence of new light particles.

\section{Beams, target and shielding}
\label{sec:Beams}
Two stages of operation are envisaged. In the first stage, lasting for about 5 years, the kinetic energy of the beam proton will be  $T_p=423$\,MeV, with the average current of   300\,$\mu$A. A rapid-cycling synchrotron, which is considered  as a second stage of the project, will have the proton energy $T_p=1.3$\,GeV. The average beam current is planned to be   100\,$\mu$A. The number of protons on target (POT) $N_\text{POT}=1.8\times 10^{22}$  per year is expected to be delivered with the proton energy of 423 MeV and an exposure of  $0.61\times 10^{22}$ POT is estimated  with  1.3\,GeV protons,  accepting the annual operation time of $10^7$\,s in both cases.  

A 2 mm diameter proton beam   hits the target,  a  graphite cylinder of a radius $r_\text{t}=10$\,cm and a  length $l_\text{t}=120$\,cm. Although pion production cross-sections are larger for heavier targets, a higher level of stopping power  in heavy targets (W, Hg) leads to protons rapidly slowing down below the pion production threshold. For low energies of protons considered in this work, a long graphite target used in the beam-dump mode provides a higher yield of pions compared to heavy targets. The graphite target is equipped with a cooling system and surrounded by a thick layer of passive shielding.     
The shielding is 
made of concrete. 
The form of the shield is a sphere of radius $R=7$\,m. The target is placed in its center. The particle detectors are located in the downstream free area. The thickness of the shielding is enough to ensure its radiation safety provided by the low energy of the beam protons, see e.g.~\cite{Chen:2024fzk}.    

\begin{figure}[!htb]
    \centering
    \includegraphics[width=\linewidth]{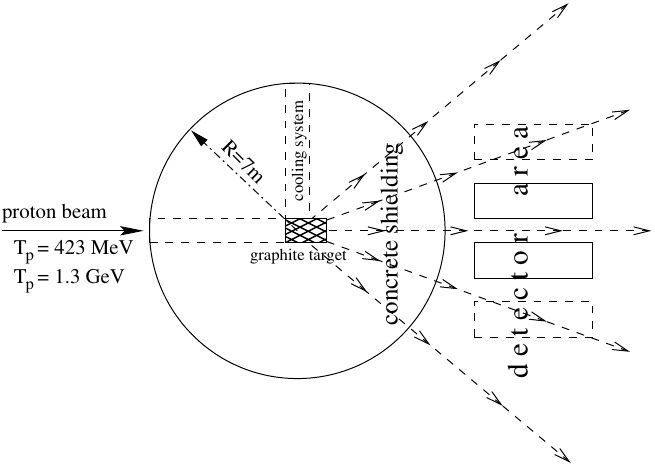}
    \caption{A sketch of the TiMoFey beam dump facility.}
    \label{fig:scheme}
\end{figure}

The scheme of the new experimental facility is shown in Fig.\,\ref{fig:scheme}. 

The incident protons in the target material produce secondary particles. 
To describe this process, simulations with Geant 4.11.2.2~\cite{GEANT4:2002zbu} have been performed. The standard QGSP\_BERT~\cite{Wright:2015xia} physics model is used to estimate the pion production on the target. This is a preferred ``physics list'' to simulate interactions of incident hadrons with $T<3$ GeV and provides the best agreement on pion-production with the thin-target data (e.g. see studies in~\cite{PhysRevD.106.032003}).

The simulations include interactions of one million protons with the target.
Table~\ref{tab:geant4}
\begin{table}[!htb]
    \centering
    \begin{tabular}{|c|c|c||c|c|c|} 
        \hline
        $T_p$,\,MeV & 423 & 1300 & $T_p$,\,MeV & 423 & 1300\\
        \hline
        $\la T_{\pi^+}\ra$,\,MeV & 63 & 330 & $\la T_{\pi^-}\ra$,\,MeV & 60 & 266\\
        \hline
        $n_0$ & $8.6\cdot 10^{-3}$ & 0.35 & $f_0$ & 0.28 & 0.35\\
        \hline
        $n_+$ & $10\cdot 10^{-3}$ & 0.54 & $f_+$ & 0.28 & 0.33\\
        \hline
        $n_-$ & $1.7\cdot 10^{-3}$ & 0.14 & $f_-$ & 0.25 & 0.29\\
        \hline
    \end{tabular}
    \caption{Results of Geant4 simulations for charged $\pi^\pm$ and neutral $\pi^0$ pion productions per proton on the graphite target for two possible kinetic energies of the TiMoFey proton beam $T_p$.}
    \label{tab:geant4}
\end{table}
\begin{table}[!htb]
\begin{tabular}{|c|c|c|c|}
\hline
Model       & $n_{\eta}$ & $f_{\eta}$ & $\la T_{\eta}\ra$,\,MeV \\ \hline
QGSP\_BIC   & $3.4\cdot 10^{-3}$     & 0.19       & 156               \\ \hline
QBBC        & $2.3\cdot 10^{-3}$     & 0.2        & 148               \\ \hline
QGSP\_INCLXX & $0.8\cdot 10^{-3}$     & 0.14       & 154               \\ \hline
\end{tabular}
\caption{Results of Geant4 simulations for $\eta$-meson productions per proton on graphite target for $T_p = 1300\MeV$ and for different hadronic models.}
\label{tab:eta_prod}
\end{table}
presents the results of the simulations for both possible values of the proton beam kinetic energy $T_p$. For each proton all inelastic scatterings are accounted for, though mostly the first scatterings contribute to the meson production. For pions with charge $i$, $n_i$ is the number of pions produced on the target per one proton of the beam, $f_i$ is the fraction of ``forward" pions with 3-momentum polar angle obeying $\theta\leq\theta_\text{bin}$, where $\cos\theta_\text{bin}=0.8$. Unlike the pion production case, a different set of Geant4 physics lists-- QGSP\_BIC~\cite{Folger:2004zma}, QBBC~\cite{Ivantchenko:2011qbbc}, and QGSP\_INCLXX~\cite{Boudard:2012wc}--is used for simulation of $\eta-$meson production. Table\,\ref{tab:eta_prod} contains  information on $\eta$-meson yields per proton, which are allowed kinematically for $T_p=1.3$\,GeV.  These models show somewhat different $\eta$-meson yields. However, there are no reliable experimental data available to clearly favor one model over another. 

The neutral pions and $\eta$-mesons decay instantly, so their kinematic characteristics can be obtained directly from simulations of the proton scatterings, see Tabs.\,\ref{tab:geant4} and \ref{tab:eta_prod}. 
The charged pions are produced on the target with mean kinetic energies $\la T_{\pi^\pm}\ra$ listed in Tab.\,\ref{tab:geant4}. Then, while passing through the graphite target, they lose energy $E_\pi$ via electromagnetic interactions with target atoms' electrons. This process is described by Bethe equation for the linear stopping power~\cite{Bethe:1930ku,RevModPhys.9.245}
\begin{equation} \label{eq:dEdx}
\begin{split}
    \left<-\frac{\dd E_\pi}{\dd l}\right>=&\\=\rho K\frac{Z}{A\beta_\pi^2}&\left(\ln\left(\frac{2 m_e \beta_\pi^2\gamma_\pi^2}{I}\right)-\beta_\pi^2-\frac{\delta(\beta_\pi\gamma_\pi)}{2} \right),
\end{split}
\end{equation}
where $l$ is the distance passed in the target. The values of graphite density $\rho$, constant $K$, atomic number $Z$ and atomic mass $A$, as well as mean excitation energy $I$ for graphite and electron mass $m_e$ are listed for convenience in Tab.\,\ref{tab:enloss}.
\begin{table}[!htb]
    \centering
    \begin{tabular}{|c|c||c|c|}
        \hline
        $Z$ & 6 & $\delta_0$ & 0.14 \\
        \hline
        $A,$\,g\,mol$^{-1}$ & 12.0 & $x_0$ & 0.0480 \\
        \hline
        $\rho,$\,g\,cm$^{-3}$ & 1.7 & $x_1$ & 2.54\\
        \hline
        $K$,\,MeV\,cm$^2$\,mol$^{-1}$ & 0.307 & $a$ & 0.208 \\
        \hline
        $m_e,$\,MeV & 0.511 & $k$ & 2.95 \\
        \hline
        $I,$\,eV & 78.0 & $\overline{C}$ & 3.16\\
        \hline
    \end{tabular}
    \caption{Parameter values used in~\eqref{eq:dEdx},\,\eqref{eq:deneff} for the graphite target \cite{ParticleDataGroup:2024cfk,Sternheimer:1983mb}.}
    \label{tab:enloss}
\end{table}
Relativistic kinematic variables of charged pions $\beta_\pi$ and $\gamma_\pi$ are defined as usual, giving
\begin{equation}
    \beta_\pi^2=\frac{E_\pi^2-m^2_{\pi^\pm}}{E_\pi^2}, \quad \beta_\pi^2\gamma_\pi^2=\frac{E_\pi^2-m^2_{\pi^\pm}}{m^2_{\pi^\pm}}.
\end{equation}
The variable $x=\log_{10}(\beta\gamma)$ is introduced to estimate the density-effect correction for conductors~\cite{Sternheimer:1983mb}
\begin{equation} \label{eq:deneff}
\begin{split}
    \delta&\left(\beta\gamma\right)=\\&=
	\begin{cases}
		\delta_0 10^{2(x-x_0)}, &x<x_0, \\
        2(\ln10)x-\overline{C}+a\left(x_1-x\right)^k, &x_0\leq x< x_1,
	\end{cases}
\end{split}
\end{equation}
the values of the fit parameters for graphite are listed Tab.\,\ref{tab:enloss}.
Using \eqref{eq:dEdx}--\eqref{eq:deneff}, we numerically integrate the inverse stopping power $1/\left(dE_\pi/dl\right)$ over pion energy $E_\pi$ from the initial pion energy $E_{i, \pi^\pm}=\la T_{\pi^\pm} \ra+m_\pi$ to the lower energy $E_{\pi}$ and therefore obtain the distance passed by the charged pion in the target $l(E_\pi)$ as a function of the charged pion energy $E_\pi$. Then we numerically find the inverse function $E_\pi(l)$ to be used below to estimate the number of signal events. It is worth mentioning that the hadronic interaction length always exceeds that of the ionization length we described above. So, most of the pions effectively stop and decay into lepton pairs, thus avoiding strong interactions.

\section{Downstream detectors}
\label{sec:Detectors}

The detector area is sufficiently large to host several experiments, suggesting the TiMoFey facility as a Center of Collective Use. 

Searches for new feebly interacting particles may be performed exploiting two generic signatures: A) appearance of a pair of SM particles from decay of a new particle inside a detector; B)  energy deposit inside a detector produced by the elastic scattering of the new particle.  
Then it is natural to place two different detectors aimed at hunting for the two types of new particles exhibiting the two different signatures.

\subsection{Detector A: Searching for pairs of particles}
Such a  detector of about  $L_{\det}=2$\,m length and cross-sectional area of~1\,m$^2$ surrounds the decay volume  placed right behind the shielding, at a distance of about $L_{\text{dump}}\simeq R=7$\,m from the pion production target. As shown in Sec.\,\ref{sec:ALP}, photons from decays of axion-like particles  have energies $10 - 300$\,MeV and $10- 800$\,MeV in the case of a 423~MeV and 1.3~GeV  proton beam, respectively. In order to search for ALPs at these proton energies in the beam-dump mode, a dedicated   $4\pi$ photon detector system capable of measuring the positions, energies, and times of medium energy photons with good resolution and high efficiency should be exploited. The reconstruction of the direction  of photons with energies larger than 100 MeV is also desirable. To meet these requirements such a  detector system  should consist of two sections:  a fine grained preradiator in which photons from ALP decays are converted,  followed by an electromagnetic  calorimeter. Layers of plastic scintillators are also to be installed before the preradiator to detect/veto charged particles from the decay volume. The preradiator has a total thickness of about three radiation lengths and is composed of thin layers of lead and segmented plastic scintillators, yielding a single-photon conversion efficiency of about 90\% and providing a two-photon detection efficiency $\sim 80$\%. Its function is to measure the position and direction (for $E_{\gamma} \geq 100$~MeV) of photons also contributing to the achievement  of sufficiently high energy resolution  by measuring the deposited energy  $e^+e^-$-pairs. The calorimeter located behind the preradiator can be made up  of “shashlyk” modules comprised of a stack of square tiles with alternating layers of lead and plastic scintillator read out by wavelength shifting  fibers. This type of calorimeter is expected to have an energy resolution of  $\sigma_E/E \simeq 2.7\%/\sqrt{E(\rm GeV)}$~\cite{Atoian:2007up} that corresponds to $8-9$\% for 100 MeV photons. The detector option  based only  on a calorimeter with smaller modules in cross-section, without the preradiator, can also be studied.  Both configurations can be used for detection of electron-positron pairs from ALP decays, but a large volume TPC surrounded by an electromagnetic calorimeter  is better suited for this purpose and will also be considered. 

\subsection{Detector B: Searching for hits}
\label{sec:detB}
A dedicated search  for millicharged particles (MCPs) can be carried out by several experimental methods. For example, it could be a detector composed of two or three layers each is an array of optically isolated plastic  scintillator bars~\cite{Ball:2020dnx}. The scintillator light produced charged particles  in the bars is detected by photomultiplier  tubes. The bars  in the layers are aligned with the direction of the target.  MCPs produced in the beam dump would pass through all layers and their signal is a  coincidence of  single-electron signals produced  in   these layers of the detector. 

Another approach is to search for separated  double  hit  events  from recoil electrons  produced by MCPs in the detector volume. These events, if produced by  MCPs, should be  aligned with the upstream meson production target. Liquid argon (LAr) detectors~\cite{Harnik:2019zee} and highly granular scintillator detectors~\cite{Blondel:2017orl,Kudenko:2025dlg} are well suited to search for MCPs using this method. In this case, the detector must have the electron detection threshold   as low as possible  to maximize the sensitivity to MCPs. To obtain the threshold of about 1~keV, a $1~{\rm m}\times 1~{\rm m}\times 2~{\rm m}$ volume is subdivided into 10 layers each containing 400 ($20\times 20$) $5~{\rm cm}\times 5~{\rm cm}\times 20~{\rm cm}$ scintillator detectors.  Each detector is optically coupled to a PMT. In plastic scintillator a minimum-ionizing particle produces about $10^4$ photons per MeV that corresponds to about 10 photons per keV. Liquid scintillators, like Linear AlkylBenzene~(LAB) based liquid scintillator, which is used in  Daya Bay and JUNO,  emit  $(1.2-1.3)\times 10^4$ scintillation photons/MeV (see, for example refs.~\cite{doi:10.1021/j100550a010, Biller:2020uoi}). Using a reflector with high reflectivity one can expect, that 80\% of photons,  produced in the detector by a recoil electron, hit the PMT photocathode, which has a common quantum efficiency of about  35\%. As a result, the light yield of about 3.5 p.e./keV can be obtained.
For   a typical single PMT photoelectron resolution of $\leq 30$\% and a detection threshold of 0.3~p.e. we obtain the  detection efficiency of $\geq 95$\% for a 1 keV recoil electron.   



\section{Axion-like particles}
\label{sec:ALP}
In this Section, we discuss the sensitivity of TiMoFey to models with light
axion-like particles (ALPs), pseudoscalars, singlet with respect to the SM gauge group. As an inherent ingredient, ALPs are present in a number of SM extensions, see, e.g.~\cite{Bauer:2017ris,Irastorza:2018dyq} and references therein. The most generic Lagrangian of this class of models contains a whole bunch of operators
describing non-renormalizable interactions of ALP with SM gauge fields and
fermions, the pattern predicted in supersymmetric extensions\,\cite{Brignole:1999gf,Gorbunov:2000th}. Here, for concreteness, we consider ALP, denoted as $a$, which interacts only with
the SM gauge fields, so the SM Lagrangian at low energies is extended by the ALP-containing part, 
\begin{equation}
  \label{eq:1.1}
  \begin{split}
  {\cal L} _{ALP} = \frac{1}{2}(\partial_\mu a)^2 - \frac{m_a^2}{2}a^2 +
  c_{GG}\frac{\alpha_s}{4\pi} \frac{a}{f} G_{\mu\nu}\tilde{G}^{\mu\nu} \\ +\,\,
  c_{WW}\frac{\alpha_2}{4\pi} \frac{a}{f} W^a_{\mu\nu}\tilde{W}^{a\mu\nu} +
  c_{BB}\frac{\alpha_Y}{4\pi} \frac{a}{f} B_{\mu\nu}\tilde{B}^{\mu\nu}.
  \end{split}
\end{equation}
Here we introduce analogues of the fine-structure constant $\alpha=\frac{e^2}{4\pi}$ for all the SM gauge interactions $SU(3)_c\times SU(2)_W\times U(1)_Y$, i.e.  
$\alpha_s = \frac{g_s^2}{4\pi}$, $\alpha_2 = \frac{g^2}{4\pi}$, $\alpha_Y
= \frac{g^{\prime 2}}{4\pi}$, and we define $f_a \equiv \frac{f}{2c_{GG}}$, where $f$ stands for the higher energy scale where the non-renormalizable Lagrangian \eqref{eq:1.1} must be completed.
Below we consider two specific benchmark scenarios: 1) gluon dominance, defined
by $c_{GG}\ne 0$ and $c_{WW}=c_{BB}=0$ (benchmark BC11 in Ref.\cite{Antel:2023hkf})  and 2) democracy, which corresponds
to the case $c_{GG}=c_{WW}=c_{BB}$. We consider the light ALPs, with
$m_a\lsim 1$\,GeV, which is relevant for  
TiMoFey proton energies. 

Interactions~\eqref{eq:1.1} give rise to light ALP production through mixing
with neutral pseudoscalar mesons. Corresponding interaction terms can be
obtained by making a chiral rotation of the light quarks
fields~\cite{Georgi:1986df} which swaps the ALP couplings to gluons for those to quarks. The phenomenology of light ALPs can be addressed within different approaches, see, e.g.~\cite{Aloni:2018vki, Bauer:2020jbp, Bauer:2021wjo, Jerhot:2022chi, DallaValleGarcia:2023xhh, Ovchynnikov:2025gpx}. 
Here we use the results of Ref.\,\cite{Aloni:2018vki} to describe the decays and production of ALP.

The differential yield of ALP produced via mixing with neutral pions and $\eta$ mesons can be written~\cite{Jerhot:2022chi} as
\begin{equation}
\label{eq:1.2}
\frac{d^2N_a}{d\theta_a dE_a} = \sum_{P=\pi^0,\eta}|\theta_{aP}|^2\,\frac{d^2N_{P}}{d\theta_{P}dE_{P}}\bigg|_{E_{P}=E_a,\,\,\theta_{P}=\theta_a}\,,
\end{equation}
where for the effective mixing angles we take~\cite{Aloni:2018vki}
\begin{equation}
\label{eq:1.3}
  \theta_{a\pi^0} =
  \frac{1}{2}\delta_I\frac{m_a^2}{m_a^2-m^2_\pi}\frac{f_\pi}{f_a}\,,\;\;
  \theta_{a\eta} =
  \frac{1}{\sqrt{6}}\frac{m_a^2-m^2_\pi/2}{m_a^2-m^2_\eta}\frac{f_\pi}{f_a}\,
\end{equation}
with isospin factor $\delta_I = \frac{m_d-m_u}{m_d+m_u}\approx \frac{1}{3}$.  

The dominant decay mode of ALPs of masses $m_a< 3m_\pi$ is $a\to\gamma\gamma$. The corresponding
decay width is given by 
\begin{equation}
\label{eq:1.4}
  \Gamma(a\to\gamma\gamma) = \frac{\alpha^2m_a^3}{256\pi^3f_a^2}|c_{\gamma\gamma}|^2\,,
\end{equation}
where the axion-photon effective coupling reads~\cite{Aloni:2018vki} 
\begin{equation}
\label{eq:1.5}
  \begin{split}
    c_{\gamma\gamma} \approx \frac{c_{WW}+c_{BB}}{c_{GG}} -
      1.92 + \frac{1}{3}\frac{m_a^2}{m_a^2-m_\pi^2} \\ +
      \,\,\frac{4}{9}\frac{2m_a^2-m_\pi^2}{m_a^2-m_\eta^2}
      +\frac{7}{9}\frac{m_a^2-2m_\pi^2}{m_a^2-m_{\eta^\prime}^2}\,.
  \end{split}
\end{equation}
For heavier ALP other decay modes, such as $a\to 3\pi$ and $a\to
\gamma\pi\pi$, open up\footnote{CP conservation forbids ALP decays into a pair of pseudoscalar mesons.}, and  they are accounted for (following Ref.\,\cite{Aloni:2018vki})  when calculating the
total decay width of the ALP to estimate the decay length and the branching ratio of $a\to\gamma\gamma$.

We suggest the ALP decay into a pair of photons inside the Detector A 
as the main ALP signature in the TiMoFey  experiment. 
The expected number of photon pairs is given by 
\begin{equation}
\label{eq:1.6}
N \!=\! N_{\text{POT}}\,\epsilon_{\det}\!\!\!\int\!\!\! d\theta_a dE_a \frac{d^2N_a}{d\theta_a dE_a}\,{\cal P}_{\det}\,{\rm Br}(a\!\to\!\gamma\gamma),
\end{equation}
where $N_{\text{POT}}$ corresponds to 5~years of operation  and the probability of ALP decay inside the detector of length $L_{\det}=2$\,m reads
\begin{equation}
\label{eq:prob-decay}
{\cal P}_{\det} = {\rm e}^{-\frac{L_{\text{dump}}}{\lambda_{a}}}\,\left(1 - {\rm e}^{-\frac{L_{\det}}{\lambda_{a}}}\right)
\end{equation}
with ALP decay length $\lambda_{a}=\gamma_a\beta_a c\tau_a$. The differential yield (per a POT) of neutral pions entering~\eqref{eq:1.2} is taken from Geant4 simulations. The production in mixing with the $\eta$ meson is relevant for $T_p=1.3$~GeV and is almost always subdominant. The corresponding yield of $\eta$ mesons is conservatively estimated using their average number produced with $\cos{\theta}>0.8$ and the average kinetic energy $\langle T_{\eta}\rangle$ presented in Table~\ref{tab:eta_prod} for the QGSP\_INCLXX hadronic model. Angular integration in~\eqref{eq:1.6} goes over all directions toward the detector volume. And we include the detection efficiency factor $\epsilon_\text{det}=0.8$. 

\begin{figure}[!htb]
  \centerline{\includegraphics[width=8.5cm]{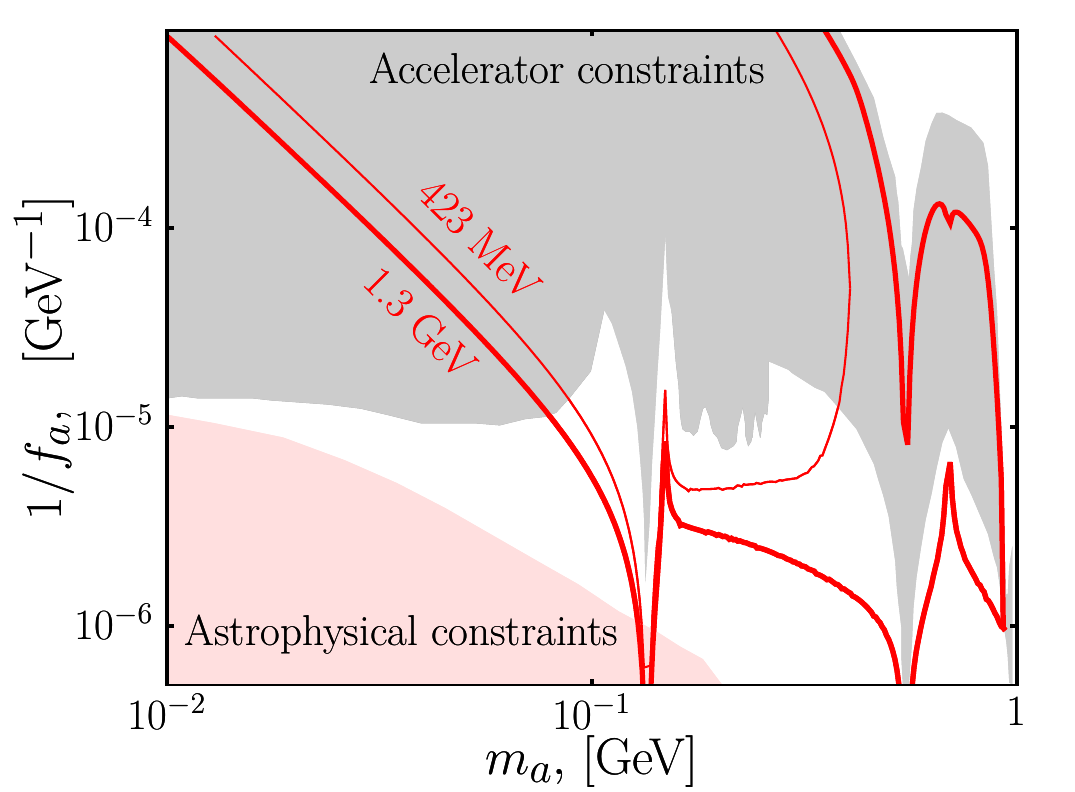}}
\caption{The 95\% CL expected sensitivity of TiMoFey (5 operation years in each stage) to ALP decaying into a pair of photons in the  gluon dominance scenario for proton kinetic energies 423\,MeV (thin red solid line) and 1300\,MeV (thick red solid line)
  in comparison with existing bounds (shaded areas). \label{fig_ax_1}}  
\end{figure}

\begin{figure}[!htb]
  \centerline{\includegraphics[width=8.5cm]{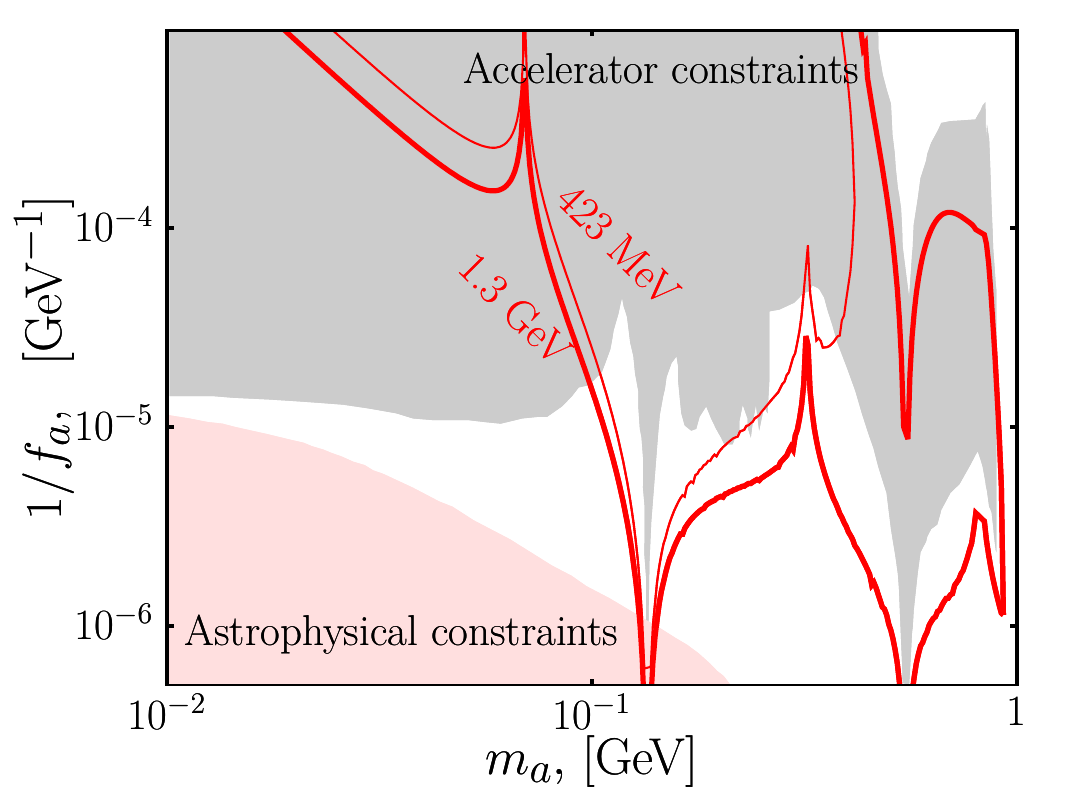}}
  \caption{The same as in Fig.~\ref{fig_ax_1} but in the democracy 
    scenario for ALP couplings. \label{fig_ax_2}} 
\end{figure}

To estimate the TiMoFey sensitivity to the ALP models, we assume the background-free case, and ask the number of signal events to be less than 3.84 which corresponds to 95\% CL for the Poisson statistics. In Figs.~\ref{fig_ax_1} and \ref{fig_ax_2} 
we show the expected sensitivity at~95~\% CL of TiMoFey to light ALPs for gluon dominance and democracy scenarios, respectively. The corresponding sensitivity contours are presented on the $(m_a,f_a)$ plane by red solid thin ($T_p=423$\,MeV) and thick ($T_p=1.3$\,GeV) lines in comparison with other existing experimental constraints (shaded regions) from kaon decays (NA62~\cite{NA62:2020xlg} and E949~\cite{BNL-E949:2009dza}), beam dump searches (CHARM~\cite{CHARM:1985anb} and NuCal~\cite{Blumlein:1990ay,Blumlein:1991xh}) recalculated in Ref.\,\cite{Jerhot:2022chi}. 
The prominent features in the expected sensitivity plots are related to resonance-like dependence of couplings in~\eqref{eq:1.3} and~\eqref{eq:1.5}. Weaker sensitivity at $m_a\lsim m_\pi$ in the democracy scenario as compared to the case of gluon dominance is attributed to a considerable numerical cancellation in the value of effective coupling constant $c_{\gamma\gamma}$ in the former case, see~\eqref{eq:1.5}.

Here we present the sensitivities to $\gamma\gamma$ final state only. However, the decay modes $a\to 3\pi$ and $a\to \gamma \pi\pi$ can also be promising for the case of proton beam energy $T_p=1.3$~GeV.

\section{Leptophobic \texorpdfstring{$B$}{B}-boson}
\label{sec:B-boson}

In this Section, we study expected TiMoFey sensitivity to the visible decays of the so-called leptophobic $B$-boson, which is a hypothetical vector massive particle, associated with the gauge field of the baryonic $U(1)_B$ group. Its interactions with quarks and leptons are described by the Lagrangian~\cite{Tulin:2014tya}
\begin{equation}
    \mathcal{L}_\text{int}=\left(\frac{g_B}{3}+\epsilon Q_q e\right) \bar{q} \gamma^\mu q B_\mu - \epsilon e \bar{l}\gamma^\mu l B_\mu.
\end{equation}
Here $g_B$ is the baryonic gauge coupling, $e$ is the proton electric charge, $Q_q e$ is the quark electric charge. The interactions of the $B$-boson with leptons appear due to its mixing with the SM photon via quark loops. The corresponding coupling is proportional to the parameter $\epsilon\ll 1$, which is equivalent to the dark photon ($\gamma^\prime$) kinetic mixing. In principle, the model contains three independent parameters: gauge coupling, kinetic mixing, and boson mass, $(g_B, \epsilon, m_B)$. For simplicity, we accept the following widely used relationship between $\epsilon$ and $g_B$ typical for one-loop radiation corrections
\begin{equation}
    \epsilon=\frac{e g_B}{\left(4\pi\right)^2}.
\end{equation}

In this study, we consider three possible $B$-boson production channels. Two of them are associated with the secondary mesons produced by the protons incident on the target, and the last one is associated with the direct collisions of protons with the target. 

The total number of pions with charge $i$ ($\eta$-mesons) produced at the first inelastic proton scatterings is 
\begin{equation}
    N_{\pi^i (\eta)}=\frac{I_pt}{e}n_{i(\eta)}f_{i(\eta)},
\end{equation}
where $I_p$ is the proton electric current and $t$ is the operation time, see Sec.\,\ref{sec:Beams}. The mean kinetic energies of the charged pions produced on the target $\la T_{\pi^\pm}\ra$, the pion ($\eta$-meson) yield $n_{i(\eta)}$ and the ratio of pions ($\eta$-mesons) moving in the forward direction $f_{i(\eta)}$ are presented in Tab.\,\ref{tab:geant4}\,(\ref{tab:eta_prod}). In order to {\it conservatively} estimate the sensitivity of TiMoFey to model parameters of $B$-bosons produced in $\eta$-meson decays, below we use the results of simulation of $\eta$-meson production, performed with the hadronic model QGSP\_INCLXX and listed in Tab.\,\ref{tab:eta_prod}. 

\subsection{\texorpdfstring{$B$}{B}-boson production in neutral meson decays}

For $B$-bosons lighter than the neutral pion, the dominant production channel is the decay $\pi^0\rightarrow B\gamma$. The branching ratio of neutral pion into $B$-boson and photon reads\,\cite{Tulin:2014tya}
\begin{equation}
    \text{Br}\left(\pi^0\rightarrow B\gamma\right) \simeq \frac{2\alpha_B}{\alpha_\text{em}} \left(1-\frac{m^2_B}{m^2_{\pi^0}}\right)^3 |F_\omega(m_B^2)|^2. 
\end{equation}
Hereafter $\alpha_B=g^2_B/4\pi$ is the baryonic fine structure constant and the VMD form factor (we adopt a  hypothesis of the vector meson dominance) is given by 
\begin{equation} \label{eq:VMDff}
    F_\omega(m_B^2) = \frac{m^2_\omega}{m^2_\omega-m^2_B},
\end{equation}
where we have neglected the decay width contribution of the Breit-Wigner type to the denominator, since we always consider $m_B \ll m_\omega$. 
It appears due to the mixing of $B$-boson with $\omega$-meson that has mass $m_\omega=783\,\text{MeV}$ and the same quantum numbers, $J^{PC}=1^{--}$, as $B$-boson~\cite{ParticleDataGroup:2024cfk}.

The impact of $\eta$-meson decay $\eta \rightarrow B\gamma$ into $B$-boson production becomes significant for masses $150\,\text{MeV}\lesssim m_B \lesssim 550\,\text{MeV}$. In the framework of the VMD hypothesis, the corresponding branching ratio can be estimated using~\cite{Tulin:2014tya}
\begin{equation}
    \frac{\text{Br}\left(\eta\rightarrow B\gamma\right)}{\text{Br}\left(\eta\rightarrow \gamma\gamma\right)} \simeq \frac{2\alpha_B}{\alpha_\text{em}}\left(1-\frac{m^2_B}{m^2_\eta}\right)^3 |F_{\omega\phi}(m^2_B)|^2.
\end{equation}
Here the SM branching ratio of $\eta$-meson to a photon pair is $\text{Br}\left(\eta\rightarrow \gamma\gamma\right)=0.394$~\cite{ParticleDataGroup:2024cfk} and a more complicated VMD form factor 
\begin{equation}
    F_{\omega\phi}(m^2_B)=a_\omega F_\omega(m^2_B)+ a_\phi F_\phi(m^2_B)
\end{equation}
arises due to the mixing of $B$-boson not only with $\omega$-meson, but also with $\phi$-meson with mass $m_\phi=1019\,\text{MeV}$ and $J^{PC}=1^{--}$~\cite{ParticleDataGroup:2024cfk}, and hence depends on the auxiliary VMD form factors, originating from the propagators of corresponding mesons, $F_\omega(m^2_B)$ \eqref{eq:VMDff} and $F_\phi(m^2_B)$. The latter can be obtained from \eqref{eq:VMDff} by changing $m_\omega \rightarrow m_\phi$.
The coefficients
\begin{equation}
    a_\omega\equiv\frac{1}{3}\frac{c_\theta-\sqrt{2}s_\theta}{c_\theta-2\sqrt{2}s_\theta}, \quad a_\phi\equiv\frac{1}{3}\frac{2c_\theta+\sqrt{2}s_\theta}{c_\theta-2\sqrt{2}s_\theta}
\end{equation}
depend explicitly on the trigonometric functions $s_\theta\equiv \sin \theta_m$, $c_\theta\equiv \cos \theta_m$ of the $\eta-\eta^\prime$ mixing angle $\theta_m\simeq -19.5^\circ$.
\subsection{\texorpdfstring{$B$}{B}-boson production in \texorpdfstring{$\pi N$}{piN}-collisions} \label{sec:piN}
For $B$-bosons of masses $150\,\text{MeV}\lesssim m_B\lesssim 300$\,MeV an alternative source becomes important: production by charged pion scattering off nucleus, $\pi^\pm N\rightarrow B N$. 

To describe the $B$-boson production in $\pi N$-collisions we adopt the leading order ChPT Lagrangian for pion-nucleon and pion-nucleon-dark photon interactions\,\cite{Curtin:2023bcf}. However, we change the sign in the definition of $u_\mu\equiv i u^\dagger (\nabla_\mu U) u^\dagger$ in accordance with several works specializing in Heavy Baryon ChPT, see for instance~\cite{Hemmert:1997ye,Fettes:2000gb,Rijneveen:2021bfw}. In the leading order this flips the sign of the axial coupling $g_A\rightarrow-g_A$ and does not affect the squared matrix elements of the considered processes at all. The terms, relevant for $B$-boson production by pions, look as
\begin{equation} \label{eq:chpt-lagr}
\begin{split}
    \mathcal{L}_\text{int}&= (g_B+\epsilon e)B_\mu \bar{p}\gamma^\mu p + g_B B_\mu \bar{n}\gamma^\mu n +\\
    &+ i\epsilon e B_\mu \left(\pi^-\partial^\mu \pi^+-\pi^+\partial^\mu \pi^-\right)-\\
    &- \frac{g_A}{f_\pi\sqrt{2}}\left(\bar{p}\gamma_\mu\gamma_5 n\,\left(\partial^\mu -i\epsilon e B^\mu \right)\pi^+ +\right.\\
    &\left.+ \bar{n}\gamma_\mu\gamma_5p\,\left(\partial^\mu+i\epsilon e B^\mu\right)\pi^-\right),
\end{split}
\end{equation}
with the axial coupling $g_A=1.27$. 

\begin{figure}[!htb]
	\begin{center}
		\begin{subfigure}{0.23\textwidth}
			\centering
                \includegraphics[width=\textwidth]{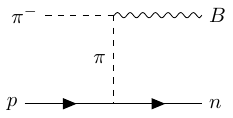}
			\caption{}
		\end{subfigure}
        \hfill
		\begin{subfigure}{0.23\textwidth}
			\centering
			\includegraphics[width=0.93\textwidth]{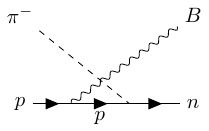}
			\caption{}
		\end{subfigure}
        \begin{subfigure}{0.23\textwidth}
			\centering
			\includegraphics[width=\textwidth]{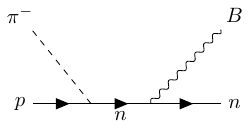}
			\caption{}
		\end{subfigure}
        \hfill
        \begin{subfigure}{0.23\textwidth}
			\centering
			\includegraphics[width=0.95\textwidth]{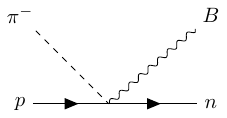}
			\caption{}
		\end{subfigure}
	\end{center}
	\caption{Feynman diagrams for $B$-boson production in $\pi^- p$-collisions in extended Heavy Baryon ChPT: (a) $t$-channel~\eqref{eq:mint}, (b) $u$-channel~\eqref{eq:minu}, (c) $s$-channel~\eqref{eq:mins}, (d) 4-point~\eqref{eq:min4}. Figures were made with the help of the package \texttt{TikZ-Feynman}~\cite{Ellis:2016jkw}.}
        \label{fig:pipFeyn}
\end{figure}

Figure\,\ref{fig:pipFeyn}
shows Feynman diagrams for $\pi^-p\rightarrow Bn$. Another scattering process, contributing to $B$-boson production, $\pi^+n\rightarrow Bp$, is described by four analogous diagrams.

Matrix elements for the process $\pi^-(p) p(P_i)\rightarrow n(p_n)B(k)$ can be written as 
\begin{align}
    \begin{split} \label{eq:mint}
        \mathcal{M}^-_t=&-\frac{g_A \epsilon e\sqrt{2}}{f_\pi\left(t-m^2_{\pi^\pm}\right)} \left(p\cdot \epsilon^*(k)\right)\times\\\times& \bar{n}(p_n)(\hat{p}-\hat{k})\gamma_5 p(P_i),
    \end{split}\\
    \begin{split} \label{eq:minu}
        \mathcal{M}^-_u=&\frac{g_A(g_B+\epsilon e)}{f_\pi\sqrt{2}\left(u-m_N^2\right)}\times\\\times& \bar{n}(p_n)\hat{p}\gamma_5\left(\hat{P_i}-\hat{k}+m_N\right)\widehat{\epsilon^*}(k)p(P_i),
    \end{split}\\
    \begin{split} \label{eq:mins}
        \mathcal{M}^-_s=&\frac{g_Ag_B}{f_\pi\sqrt{2}\left(s-m^2_N\right)}\times\\\times&\bar{n}(p_n)\widehat{\epsilon^*}(k)\left(\hat{p}_n+\hat{k}+m_N\right)\hat{p}\gamma_5 p(P_i),
    \end{split}\\ \label{eq:min4}
    \mathcal{M}^-_4=&\frac{g_A\epsilon e}{f_\pi\sqrt{2}}\bar{n}(p_n)\widehat{\epsilon^*}(k)\gamma_5p(P_i),
\end{align}
where $p$ is the momentum of $\pi^-$, $P_i$ is the momentum of the target proton, $k$ is $B$-boson momentum and $p_n$ is the neutron momentum. 
In order to check~\eqref{eq:mint}--\eqref{eq:min4}, we have compared the amplitudes at $m_B\ll E_\pi$ with the result published in~\cite{Shin:2022ulh} for the effectively massless dark photon considered to be emitted in analogous processes inside a supernova. Our results do not agree with~\cite{Shin:2022ulh}, see also the comment on gauge invariance of the full amplitude below.

The process $\pi^+(p) n(P_i)\rightarrow p(p_p)B(k)$ (variables in brackets stand for particles momenta) is described by the following matrix elements 
\begin{align}
    \begin{split}
        \mathcal{M}^+_t=&\frac{g_A \epsilon e\sqrt{2}}{f_\pi\left(t-m^2_{\pi^\pm}\right)} \left(p\cdot \epsilon^*(k)\right)\times\\\times& \bar{p}(p_p)(\hat{p}_p-\hat{P}_i)\gamma_5 n(P_i),
    \end{split}\\
    \begin{split}
        \mathcal{M}^+_u=&\frac{g_Ag_B}{f_\pi\sqrt{2}\left(u-m_N^2\right)}\times\\\times& \bar{p}(p_p)\hat{p}\gamma_5\left(\hat{P_i}-\hat{k}+m_N\right)\widehat{\epsilon^*}(k)n(P_i),
    \end{split}\\
    \begin{split}
        \mathcal{M}^+_s=&\frac{g_A(g_B+\epsilon e)}{f_\pi\sqrt{2}\left(s-m^2_N\right)}\times\\\times&\bar{p}(p_p)\widehat{\epsilon^*}(k)\left(\hat{P}_i+\hat{p}+m_N\right)\hat{p}\gamma_5 n(P_i),
    \end{split}\\
    \mathcal{M}^+_4=&-\frac{g_A\epsilon e}{f_\pi\sqrt{2}}\bar{p}(p_p)\widehat{\epsilon^*}(k)\gamma_5n(P_i).
\end{align}
Numerically, they give an even bigger input in the final number of events due to the enhanced production of positively charged pions, see Tab.\,\ref{tab:geant4}. 

For each process we calculate the total matrix elements
\begin{equation}
    \mathcal{M}^\mp=\mathcal{M}^\mp_t+\mathcal{M}^\mp_u+\mathcal{M}^\mp_s+\mathcal{M}^\mp_4,
\end{equation}
using the \texttt{FeynCalc} package~\cite{Mertig:1990an,Shtabovenko:2016sxi,Shtabovenko:2020gxv} for \texttt{Wolfram Mathematica}. We also check the gauge invariance of the obtained result by replacing the polarization vector of leptophobic $B$-boson $\epsilon^*_\mu (k)$ with its momentum $k^\mu$ and taking the massless limit $m_B \ll E_\pi$. For both total matrix elements $\mathcal{M}^\mp$ we obtain zero upon such replacements. Most probably, Ref.\,\cite{Shin:2022ulh} contains a misprint in the expressions for the corresponding amplitudes, since they do not pass such a test, despite the fact that it is mentioned in the text.

Finally, we obtain the differential cross section for $\pi N\rightarrow BN$ processes
\begin{equation}
    \frac{\dd \sigma_\mp}{\dd t}=\frac{|\mathcal{M}^\mp|^2}{128\pi m^2_N T(T+2m_\pi)},
\end{equation}
to be integrated over the Mandelstam variable $t$ from $t(\theta_\text{bin})$ to $t(0)$. In terms of the  polar angle $\theta_\text{cm}$, corresponding to $B$-boson 3-momentum in the center-of-mass frame, this variable reads 
\begin{equation}
\begin{split}
    t(\theta_\text{cm})=&\left(E^\text{cm}_\pi-E^\text{cm}_B\right)^2-\left(|\vec{p}_\text{cm}|-|\vec{k}_\text{cm}|\right)^2 -\\&-4|\vec{p}_\text{cm}||\vec{k}_\text{cm}| \sin^2\frac{\theta_\text{cm}}{2}
\end{split}
\end{equation}
and the expressions for the energies and absolute values of momenta in the center-of-mass frame for the particles in $2\rightarrow2$ process can be found in~\cite{ParticleDataGroup:2024cfk}.

\subsection{\texorpdfstring{$B$}{B}-boson production in \texorpdfstring{$pN$}{pN}-collisions}
One of the most significant contributions to $B$-boson production comes from the $2\rightarrow 3$ process of proton-nuclei collisions $pN\rightarrow BpN$. This process is similar to elastic proton bremsstrahlung, playing an important role in searches for dark photon along with $\pi^0$- and $\eta$-mesons decays~\cite{Chen:2024fzk}. Following section~\ref{sec:piN}, we study it in the framework of ChPT with baryons using the Lagrangian~\eqref{eq:chpt-lagr}, which we extend by including the interaction of neutral pions with nucleons
\begin{equation}
    \mathcal{L}_{\text{int},\,\pi^0}=\frac{g_A}{2F}\left(\bar{n}\gamma^\mu \gamma^5 n - \bar{p}\gamma^\mu\gamma^5 p\right)\partial_\mu \pi^0.
\end{equation}
The related calculations in the framework of Heavy Baryon ChPT for dark gauge bosons production in $NN$-bremsstrahlung inside the supernovae were earlier performed in~\cite{Shin:2021bvz}.

\begin{figure}[!htb]
	\begin{center}
		\begin{subfigure}{0.27\textwidth}
			\centering
                \includegraphics[width=\textwidth]{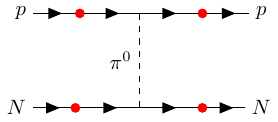}
			\caption{}
		\end{subfigure}
        \hfill
		\begin{subfigure}{0.33\textwidth}
			\centering
			\includegraphics[width=\textwidth]{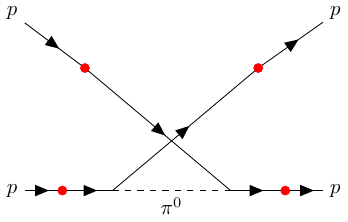}
			\caption{}
		\end{subfigure}
        \begin{subfigure}{0.33\textwidth}
			\centering
			\includegraphics[width=\textwidth]{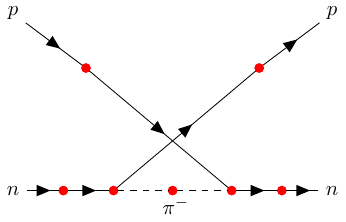}
			\caption{}
		\end{subfigure}
        \hfill
	\end{center}
	\caption{Feynman diagrams sketches for $B$-boson production in $pN$-collisions in extended Heavy Baryon ChPT: (a) $t$-channel for both $pp$- and $pn$-scatterings, (b) $u$-channel for $pp$-scattering, (c) $u$-channel for $pn$-scattering. Red dots denote various places where $B$-boson outgoing line can be inserted. Figures were made with the help of the package \texttt{TikZ-Feynman}~\cite{Ellis:2016jkw}.}
        \label{fig:pNFeyn}
\end{figure}

Figure~\ref{fig:pNFeyn}
shows the sketches of Feynman diagrams for the process $pN\rightarrow BpN$. Here we place in one figure the sketches that correspond to the same $pN$-scattering subprocesses (e.g. scattering in $t$- and $u$-channels) and mark with red dots the edges and vertices, from which $B$-boson can be emitted. In total, elastic proton bremsstrahlung $pp\rightarrow Bpp$ and hadron bremsstrahlung $pn\rightarrow Bpn$ are described by 8 and 11 Feynman diagrams, correspondingly. All matrix elements for these processes and their squares were obtained with symbolic calculations using the \texttt{CompHEP} package \cite{Pukhov:1999gg,CompHEP:2004qpa}. 

The differential cross section was integrated over the phase volume with the help of the Monte Carlo generator implemented in \texttt{CompHEP}. To account for the attenuation of the protons due to the radiation losses in the target, each differential cross section was integrated for four different incident proton energies $E_{p,j}\equiv E_p((j+1/2)\Delta l),\,j=\overline{0,3}$. Here we place the origin at the front side of the target, choose the shortest distance $l_p=\min(\lambda_\text{int}, l_\text{th})$ among the nuclear interaction length for graphite $\lambda_\text{int}=38.8\,\text{cm}$~\cite{ParticleDataGroup:2024cfk} and the position $l_\text{th}=l(E_p=E_\text{th})$ where the proton reaches the reaction threshold $E_\text{th}=m_N+m_B\left(2+{m_B}/(2m_N)\right))$, and then divide it into four equal lengths $\Delta l$. This rough discretization is dictated by high time costs and the operation of the \texttt{CompHEP} package in the manual mode. We also imposed the cut on the polar angle of $B$-boson three-momentum $\tan \theta < d/(2L_\text{dump})$, where $d=100\,\text{cm}$ is the length of the front side of the detector. Below we denote the resulting integrated cross sections of proton and hadron bremsstrahlung as $\sigma_{pp}(E_{p,j})$ and $\sigma_{pn}(E_{p,j})$.
\subsection{Sensitivity of TiMoFey to visible \texorpdfstring{$B$}{B}-boson decays}

Being feebly interacting, $B$-boson can penetrate the shielding and decay inside the detector A into a pair of SM particles, providing a distinct signature. 

The mixing with the SM photon induces the leptophobic $B$-boson decay into pairs of charged leptons $l^+l^-$, if lepton mass obeys $m_l<m_B/2$. The decay width is~\cite{Tulin:2014tya}
\begin{equation}
\label{Gam1}
\begin{split}
    \Gamma\left(B\rightarrow\right.& \left.l^+ l^-\right) =\\= & \frac{\alpha_\text{em}\epsilon^2 m_B}{3}\left(1+\frac{2m^2_l}{m^2_B}\right)\sqrt{1-\frac{4m^2_l}{m^2_B}}.
\end{split}
\end{equation}
For $B$-bosons lighter than the neutral pion, the only possible visible decay mode at tree level is $e^+e^-$.

For $B$-bosons with masses from neutral pion mass to about 500\,MeV the dominant decay mode is to neutral pion and photon with the decay width 
\begin{equation}
\label{Gam2}
\begin{split}
    \Gamma(B\rightarrow &\,\pi^0 \gamma) =\\= &\frac{\alpha_B \alpha_\text{em}m^3_B}{96\pi^3f_\pi^2}\left(1-\frac{m^2_\pi}{m^2_B}\right)^3 |F_\omega(m_B^2)|^2.
\end{split}
\end{equation}
The branching ratio of this mode always exceeds 95\% in the $B$-boson mass region 140--300\,MeV. However, for masses of $B$-boson closer to $\eta$-meson mass, the contributions of other hadronic decay modes, for instance $\pi^+\pi^-\pi^0$, also become important.  

In order to accurately estimate the total $B$-boson decay width, we use the package \texttt{DarkCast}~\cite{Ilten:2018crw}. This code is based on experimental results and, among other things, allows obtaining numerically the branching of new hypothetical vector bosons to $e^+e^-$-pair taking into account all possible hadronic decay modes. Thus we calculate the total $B$-boson decay width as
\begin{equation}
    \Gamma_\text{tot}=\frac{\Gamma\left(B\rightarrow e^+ e^-\right)}{\text{Br}\left(B\rightarrow e^+ e^-\right)}.
\end{equation}
The probability of the $B$-boson decay inside the detector $\mathcal{P}_\text{det}$ is given by eq.\,\eqref{eq:prob-decay} with $\lambda_a$ replaced by the $B$-boson decay length $\lambda_B=k_z/(m_B \Gamma_\text{tot})$.

To estimate the ratio of $B$-bosons that are produced at the distance $l$ from the front side of the target, propagate along the line making an angle not more than $\theta_\text{bin}$ with the beam axis, and are expected to pass through the front side of the detector of cross section $d\times d$ with $d=100\,\text{cm}$ (see Sec.\,\ref{sec:Detectors}), we introduce a factor of the geometric acceptance 
\begin{equation}
\label{geom-acc}
    r_s(l)=\frac{1}{2\pi (1-\cos\theta_\text{bin})}\frac{d^2}{(L_\text{dump}-l)^2},
\end{equation}
which is the ratio of the area of the detector front side and the spherical segment of the dump with angle $\theta_{\text{bin}}$. We assume that neutral pions and $\eta$-mesons decay very close to the front side of the target, so for the processes $\pi^0\rightarrow B\gamma$ and $\eta\rightarrow B\gamma$ we take $r_s$ at $l=0$. At the same time, in case of production by charged pions, we explicitly substitute the distance $l$ that a charged pion passed inside the target to the surface ratio $r_s(l)$. Since we have already cut the phase space while performing numerical integration in \texttt{CompHEP} for $pN$-channel, in this way for this process we accounted for the geometric acceptance. In addition, we assume that the detector efficiency is $\epsilon_\text{det}=0.8$.

This brings us to the number of events expected at TiMoFey in the case of $B$-bosons, produced dominantly in the neutral pions $\pi^0$ ($\eta$-mesons) decays,  
\begin{equation}
    N_{0(\eta)}=N_{\pi^0(\eta)}\text{Br}(\pi^0(\eta)\rightarrow B\gamma) \mathcal{P}_\text{det} r_s(0) \epsilon_\text{det},
\end{equation}
and for $B$-bosons, originating from $\pi N$-collisions, 
\begin{equation}
\begin{split}
    N_\mp=&\rho\frac{N_A}{A} N_{\pi^\mp} N_{p(n)}\times\\\times&\int_0^{l(E_\pi=m_\pi)}\!\!\!\!\dd l \int_{t(\theta_\text{bin})}^{t(0)} \!\!\!\!\!\dd t \frac{\dd \sigma_\mp(E(l))}{\dd t} \mathcal{P}_\text{det}r_s(l) \epsilon_\text{det},
\end{split}
\end{equation}
where the multiplier $\gamma\beta$ in the expression~\eqref{eq:prob-decay} for the probability of $B$-boson decay inside the detector $\mathcal{P}_\text{det}$ is restored in the integration over the variable $t$ and $N_{p(n)}=6$ is the number of protons (neutrons) per one target atom. Following~\cite{Curtin:2023bcf}, we use the following approximation for pion-atomic cross sections
\begin{align}
    \sigma(\pi^-C_6^{12}\rightarrow B\,B^{12}_5)&\simeq N_p \sigma(\pi^- p\rightarrow Bn),\\
    \sigma(\pi^+C_6^{12}\rightarrow B\,N^{12}_7)&\simeq N_n \sigma(\pi^+ n\rightarrow Bp).
\end{align}
Finally, the number of events expected at TiMoFey for $B$-bosons produced via proton bremsstrahlung $pp\rightarrow Bpp$ and that via the analogous process $pn\rightarrow Bpn$ are 
\begin{equation}
\begin{split}
    N_{pp(pn)}=&\rho\frac{N_A}{A}N_\text{POT}N_{p(n)}\Delta l\, \epsilon_\text{det}\times \\\times &\sum_{j=0}^3 w_j \sigma_{pp(pn)}(E_{p,j}) \mathcal{P}_\text{det}, 
\end{split}
\end{equation}
where $w_j=\{0.5, 1, 1, 0.5\}$ are the weights for trapezoidal numerical integration over the proton path in the target and here for $\mathcal{P}_\text{det}$~\eqref{eq:prob-decay} we use $\gamma\beta\equiv \la p_{B,L}\ra/m_B$ with longitudinal momentum $\la p_{B,L}\ra$ averaged over the distribution for $B$-bosons obtained with \texttt{CompHEP}.

\begin{figure}[!htb]
    \centering
    \includegraphics[width=0.98\linewidth]{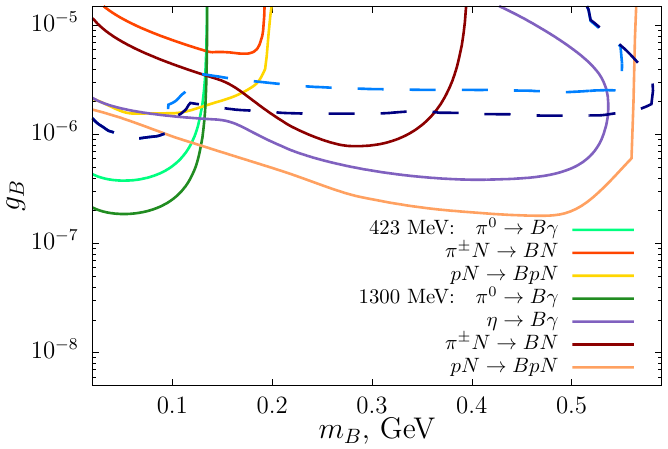}
    \caption{Solid lines: expected sensitivity of TiMoFey (5 years of operation in each stage) to visible $B$-boson decays at 95\%\,CL for the channels of neutral pion decays $\pi^0\rightarrow B\gamma$ (light green line for beam kinetic energy $T_p=423\,\text{MeV}$, dark green line for $T_p=1300\,\text{MeV}$), $\eta$-meson decays $\eta\rightarrow B\gamma$ (violet line, possible only for $T_p=1300\,\text{MeV}$), pion-nucleon collisions $\pi^\pm N\rightarrow BN$ (light red line for $T_p=423\,\text{MeV}$, dark red line for $T_p=1300\,\text{MeV}$) and proton-nucleon collisions $pN\rightarrow BpN$ (in yellow for $T_p=423\,\text{MeV}$, in orange for $T_p=1300\,\text{MeV}$). Regions in the parameter space above dashed lines are excluded at 95\%\,CL by existing (a) modern conservative experimental limits according to~\cite{NA62:2023qyn,Astier:2001ck,Tsai:2019buq,Bernardi:1985ny}, shown in light blue; (b) original strong experimental limits according to~\cite{Bergsma:1985qz,NA62:2023qyn,Astier:2001ck,Blumlein:1990ay,Blumlein:1991xh}, shown in dark blue.}
    \label{fig:Bbosonsens}
\end{figure}

In Fig.\,\ref{fig:Bbosonsens}
we outline the expected TiMoFey sensitivity at 95\%\,CL to visible decays of leptophobic $B$-boson after 5 years of TiMoFey operation in each stage. We consider four possible channels of $B$-boson production: the decays of neutral pions $\pi^0\rightarrow B\gamma$ (green lines), $\eta$-mesons $\eta\rightarrow B\gamma$ (violet line), the charged pion-nucleon collisions $\pi^\pm N\rightarrow BN$ (red lines) and the proton-nucleon collisions $pN\rightarrow BpN$ (yellow and orange lines). In the relevant cases the limits referring to the protons with kinetic energy $T_p=423\,(1300)\,\text{MeV}$ are shown in light (dark) colors. In the same figure, with the help of the package \texttt{DarkCast}~\cite{Ilten:2018crw} (see also the update~\cite{Foguel:2022ppx}), we also show the recasted by us experimental limits, originally obtained for dark photons $\gamma^\prime$ and $U(1)_{B-L}$ gauge bosons. We depict two sets of joined experimental limits with dashed lines. In light blue we combine the most sensitive conservative limits which we obtain from rescaling of the NA62~\cite{NA62:2023qyn}, NOMAD~\cite{Astier:2001ck}, NuCAL~\cite{Tsai:2019buq} and PS191~\cite{Bernardi:1985ny} data. The rescaling is based on the recent reevaluation of the pions produced in the NuCAL experiment, performed in Ref.\,\cite{Tsai:2019buq}.  Dark blue line shows the unification of the strongest original experimental limits from CHARM~\cite{Bergsma:1985qz}, NA62~\cite{NA62:2023qyn}, NOMAD~\cite{Astier:2001ck} and NuCAL~\cite{Blumlein:1990ay,Blumlein:1991xh} experiments. 

There exist other possible production channels that have not been studied in this work, but require further thorough investigation. They are the cascade processes with the production of $\Delta(1232)$-resonance like $pp\rightarrow\Delta^+p$, $pn\rightarrow\Delta^+ n$, $pn\rightarrow \Delta^0 p$ and its subsequent decay to the corresponding nucleon $\Delta\rightarrow BN$. These cascades were earlier sketched for LSND in~\cite{Batell:2009di}.   

\section{Millicharged Particles}
\label{sec:MCP}

One of the possible extensions of the Standard Model (SM) involves the existence of particles carrying a small electric charge $\varepsilon e$, where $e$ is the electron charge and $\varepsilon \ll 1$. Such millicharged particles (MCPs) naturally arise in theoretical frameworks that include additional particles with small hypercharge \cite{Foot:1990mn}, or in models featuring kinetic mixing between the photon and a hidden sector massive dark photon \cite{Holdom:1985ag}. The phenomenology of MCPs is determined by their coupling to the electromagnetic field $A_\mu$, and for the fermion MCP $\chi$ it reads
\begin{equation}
\label{A_MCPs_int}
    \mathcal{L} = \varepsilon e A_\mu \bar{\chi} \gamma_\mu \chi\,.
\end{equation}

\begin{figure}[!htb]
    \centering
    \includegraphics[width=0.5\linewidth]{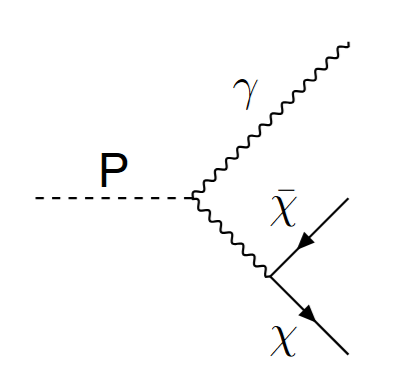}
    \caption{Feynman diagram of meson (P=$\pi$ or $\eta$) decay to photon and a pair of MCPs.}
    \label{fig:pi_gamma_chi}
\end{figure}

The interaction \eqref{A_MCPs_int} gives rise to the MCP production in neutral pion decays via the diagram in Fig.\,\ref{fig:pi_gamma_chi}. 
This decay ensures that a light meson factory is a perfect place to perform searches for MCPs with masses $m_\chi<m_\text{P}/2$. Applying Chiral Perturbation Theory (ChPT)  \cite{pich1998effective} augmented with millicharged fermion $\chi$ and interaction term \eqref{A_MCPs_int} we get for the differential decay rate of $\text{P}\to\gamma\bar\chi\chi$,  
\begin{equation}
\label{eq:pi_gamma_chi}
\begin{split}
    & \dd \Gamma = \dd \Phi \;  32\left( \frac{\varepsilon e \alpha}{2\pi f_\pi} \right)^2 \times C_{\text{P}} \times \\
    & \frac{p_\gamma q \cdot p_\gamma q \cdot p_1p_2 - p_2p_\gamma \cdot q^2 \cdot p_1p_\gamma + 2m_\chi^2\cdot p_\gamma q \cdot p_\gamma q}{q^4}
\end{split}
\end{equation}
where $p_1$ and $p_2$ are 4-momenta of $\chi$ and $\bar\chi$, $p_\gamma$ is 4-momentum of outgoing photon, $q\equiv p_1+p_2$, the pion decay constant is $f_\pi=92.4$\,MeV, $C_\eta\approx2/\sqrt{3}$, $C_\pi=1$ and d$\Phi$ is the phase space volume. We have checked that integrating this equation with $m_\chi = m_e$ and $\varepsilon=1$ gives the correct branching ratio for the neutral pion decay, $\text{Br}(\pi^0 \rightarrow \gamma e^+ e^-)=0.01$.

The pion decays are the main source of MCPs for beam dump experiments with low-energy proton beams. As the main MCP signature, we utilize a double hit in the detector volume which  previously have been exploited to impose limits on the MCP model parameters by ArgoNeuT\,\cite{ArgoNeuT:2019ckq} and suggested as a promising tool for the searches at neutrino detectors\,\cite{Gorbunov:2021jog}, e.g. SuperFGD of T2K. This signature becomes available at TiMoFey equipped with detector B. 

To estimate the TiMoFey sensitivity to MCP model parameters, we consider minimum two ionization hits of MCP inside the detector fiducial volume and treat this event as background free.   
The mean number of hits per distance $\delta x$ is described by the Rutherford differential cross section \cite{ParticleDataGroup:2020ssz, Uehling:1954wp}
\begin{equation}
\label{loss_Rutherford}
\begin{split}
   N(\delta x) & =  \frac{K}{2} \rho \frac{Z}{A} \frac{\epsilon^2}{\beta^2} \, \delta x \times \\
    & \times\int_{T_\text{min}}^{T_\text{max}} \!\!\!\!\text{d} T \,\frac{1-\beta^2T/T_\text{max} + T^2/2E^2_\chi}{T^2}\,, 
\end{split}
\end{equation}
where $T_\text{min} = 1$\,keV is the minimal electron recoil energy, $E_\chi$ is the total MCP energy, 
 $K = 0.307 \times 10^6$\,eV\,cm$^2$\,mol$^{-1}$, and the maximal energy transfer from MCP to electron $T_{\text{max}}$ comes from the scattering kinematics as
\begin{equation}
\label{Tmax}
    T_{\text{max}} = \frac{2m_e\beta^2\gamma^2}{1+\frac{2\gamma m_e}{m_\chi} + \left( \frac{m_e}{m_\chi} \right)^2}\,.
\end{equation}
The relevant values of the parameters entering eq.\,\eqref{loss_Rutherford} are presented in Sec.\,\ref{sec:detB}.

Then the probability of MCP to produce 2 hits is equal to 
\begin{equation}
    P = 1-e^{-N(L_\text{det})} - N(L_\text{det})e^{-N(L_\text{det})} \approx \frac{N^2(L_\text{det})}{2},
\end{equation}
where $L_\text{det}$ denotes the length of detector B. 

The energy and angular distributions of MCPs are determined by the corresponding distributions of their parent mesons. The energy and angular distributions of neutral pions ($\pi^0$) are calculated in Sec.\,\ref{sec:Beams}. We integrate the production rate given by Eq.\,\eqref{eq:pi_gamma_chi} using the normalized pion momentum distribution $\frac{\dd n_\pi}{\dd p \, \dd \cos \theta_\pi}$. To account for the angular distribution of the $\pi^0$ we include only pions with  $\cos\theta_\pi \in [0.8;1.0]$ and approximate the distribution over $\cos\theta_\pi$ there as flat. For $\eta$ mesons we took the mean number of mesons produced with angles $\cos\theta>0.8$ and the mean energy as in Tab.\ref{tab:eta_prod}.

\begin{figure}[!htb]
    \centering
    \includegraphics[width=0.95\linewidth]{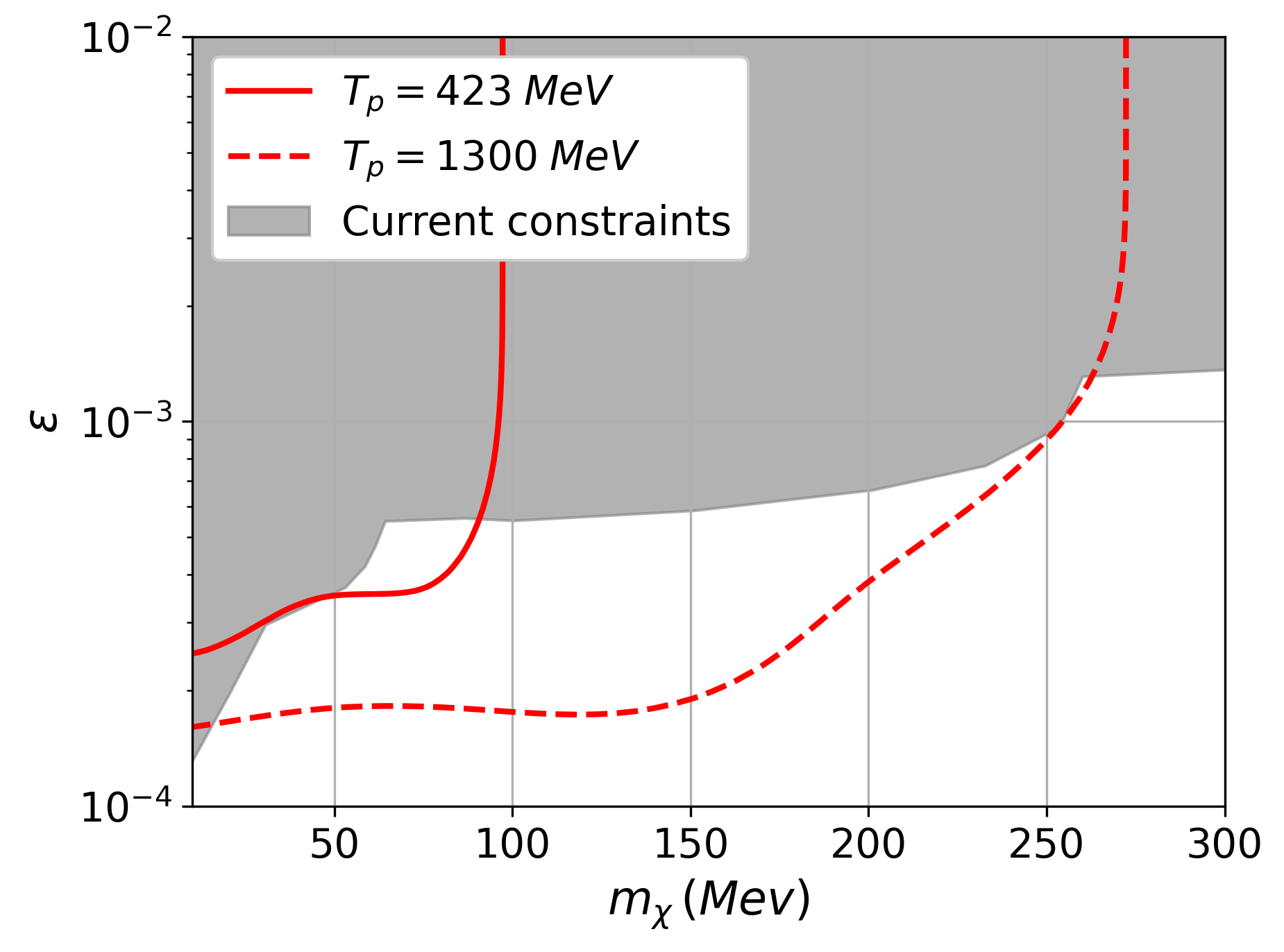}
    \caption{The regions above the green and red lines will be probed at 95\% CL after five years of operation with proton kinetic energy of 423\,MeV and 1300\,MeV correspondingly. The existing limits (shaded area) represent combined constraints of SLAC-mQ, LSND, BEBC and SENSEI \cite{Arefyeva:2022eba, PhysRevLett.122.071801, 10.21468/SciPostPhys.10.2.043,SENSEI:2023gie}.}
    \label{fig:MCPs}
\end{figure}

Another potential source of MCPs is $\pi \, N$ and $p \, N$ scattering. The methodology is analogous to that described in Sec.\,\ref{sec:B-boson} with the substitution of the B-boson by a virtual, Standard Model, photon producing MCPs through the interaction  \eqref{A_MCPs_int}. Subsequently, we consider processes $\pi^+ n \rightarrow \chi \bar{\chi}p$, $\pi^- p \rightarrow \chi \bar{\chi}n$ and $p \, n \rightarrow p \, n \, \chi \bar{\chi}$. The $pp$ scattering is suppressed in comparison with the $pn$. The expected number of signal events is given by
\begin{equation}
\begin{split}
    N_{pn}=&\rho\frac{N_A}{A} N_{p(n)} \; \epsilon_\text{det}\times\\\times&\int_0^{l\approx40\, cm} \dd l \int \frac{\dd \sigma_{pn}(E(l))}{\dd \Phi} \, \dd \Phi \, P, 
\end{split}
\end{equation}
where $\Phi$ is the phase-space volume corresponding to $\cos \theta_\chi>\cos \theta_\text{det}$,
\begin{equation}
\begin{split}
    N_\mp=&\rho\frac{N_A}{A} N_{p(n)} N_{\pi^\mp}\; \epsilon_\text{det}\times\\\times&\int_0^{l(E_\pi=m_\pi)} \dd l \int \frac{\dd \sigma_\mp(E(l))}{\dd \Phi} \, \dd \Phi \, P \; r_s(l), 
\end{split}
\end{equation}
and for the events from decays of $\pi^0$ and $\eta$, 
\begin{equation}
   N_\chi = \int_\Phi N_\text{P} \times \dd \Gamma \times P \times \dd n_\pi \times r_s(0) \times \epsilon_\text{det},
\end{equation}
where $\dd \Phi$ denotes the phase space volume of $\chi$ with angles $\cos \theta_\chi>0.8$ and $r_s(l)$ is the factor of geometric acceptance\,\eqref{geom-acc}. $\epsilon_\text{det} = 0.95^2$ is the detection efficiency, assuming a single recoil electron detection efficiency of 0.95, as presented in Sec.~\ref{sec:detB}. We ask $N_\chi>3.84$ to find the expected limit at 95\% CL which can be achieved with 5 years of TiMoFey operation. The result is shown in Fig.\,\ref{fig:MCPs}. 
Currently, the most stringent constraints on MCP models for 1\,MeV$<m_\chi <$ few 100\,MeV come from the revised data of SLAC-mQ \cite{Arefyeva:2022eba}, LSND experiment \cite{PhysRevLett.122.071801}, BEBC \cite{10.21468/SciPostPhys.10.2.043} and SENSEI \cite{SENSEI:2023gie}, they are also depicted in the plot.

The number of signal events scales as $\varepsilon^6$, since the differential decay width behaves as $\dd \Gamma \propto \varepsilon^2$, and the detection probability scales as $P \propto \varepsilon^4$. Consequently, the experimental sensitivity to the millicharge parameter scales with the total number of neutral pions as $\varepsilon_\text{min} \propto N_\pi^{-1/6}$. This relatively weak dependence justifies the use of a simplified model for the initial meson distribution in our analysis. This also indicates that, in order to enhance the sensitivity of the experiment to models involving MCPs, one should consider varying the detector and shielding materials. Different material choices can influence the interaction probability and energy loss of MCPs, potentially improving the overall detection efficiency.
~\\

\section{Conclusion}

To summarize, we considered the proposal of a new project of accelerator complex in Troitsk as multipurpose center TiMoFey providing with high intensity proton beam, which allows one to run a beam dump experiment with high statistics of hadron collisions. 
We have estimated the TiMoFey prospects in searches for new light axion-like particles, hidden vectors, associated with baryon current, and millicharged particles. In all cases the new facility can explore new regions in the model parameter spaces, potentially improving by up to an order of magnitude bound on couplings of new particles in subGeV mass range. 

This is the first paper devoted to TiMoFey project, and other models with new light particles may be studied, and the present estimates may be refined (including e.g. new production mechanisms and other signatures), once the detailed description of the detectors to be installed will become available. The main task for the TiMoFey project is measurements of neutrino coherent cross sections, which deserves a separate publication.  It is worth mentioning the present discussion of the possible third stage of TiMoFey operation when the accelerator complex is accomplished by a superconducting Linac. It will give a decent chance to increase the proton current by a factor of several with a subsequent increase of event statistics.

\begin{acknowledgments}
We thank the Nuclear Physics Division of the Physics Department of the Russian Academy of Sciences for its support of the Project and interest in the Physics case for the TiMoFey. 

EK thanks Ana Luisa Foguel for the references and for helpful discussions on recasting experimental sensitivities. 
The study of new physics phenomenology is supported by the RSF grant No. 25-12-00309. 
The work of EK on calculation of B-boson production cross sections is supported by the PhD fellowship No. 21-2-10-37-1 of the Foundation for the Advancement of Theoretical Physics and Mathematics ``BASIS". The work of DK on calculation of MCP production cross sections is supported by the Foundation for the Advancement of Theoretical Physics and Mathematics “BASIS”.
\end{acknowledgments}

\bibliography{refs}

\begin{thebibliography}{70}%
\makeatletter
\providecommand \@ifxundefined [1]{%
 \@ifx{#1\undefined}
}%
\providecommand \@ifnum [1]{%
 \ifnum #1\expandafter \@firstoftwo
 \else \expandafter \@secondoftwo
 \fi
}%
\providecommand \@ifx [1]{%
 \ifx #1\expandafter \@firstoftwo
 \else \expandafter \@secondoftwo
 \fi
}%
\providecommand \natexlab [1]{#1}%
\providecommand \enquote  [1]{``#1''}%
\providecommand \bibnamefont  [1]{#1}%
\providecommand \bibfnamefont [1]{#1}%
\providecommand \citenamefont [1]{#1}%
\providecommand \href@noop [0]{\@secondoftwo}%
\providecommand \href [0]{\begingroup \@sanitize@url \@href}%
\providecommand \@href[1]{\@@startlink{#1}\@@href}%
\providecommand \@@href[1]{\endgroup#1\@@endlink}%
\providecommand \@sanitize@url [0]{\catcode `\\12\catcode `\$12\catcode `\&12\catcode `\#12\catcode `\^12\catcode `\_12\catcode `\%12\relax}%
\providecommand \@@startlink[1]{}%
\providecommand \@@endlink[0]{}%
\providecommand \url  [0]{\begingroup\@sanitize@url \@url }%
\providecommand \@url [1]{\endgroup\@href {#1}{\urlprefix }}%
\providecommand \urlprefix  [0]{URL }%
\providecommand \Eprint [0]{\href }%
\providecommand \doibase [0]{http://dx.doi.org/}%
\providecommand \selectlanguage [0]{\@gobble}%
\providecommand \bibinfo  [0]{\@secondoftwo}%
\providecommand \bibfield  [0]{\@secondoftwo}%
\providecommand \translation [1]{[#1]}%
\providecommand \BibitemOpen [0]{}%
\providecommand \bibitemStop [0]{}%
\providecommand \bibitemNoStop [0]{.\EOS\space}%
\providecommand \EOS [0]{\spacefactor3000\relax}%
\providecommand \BibitemShut  [1]{\csname bibitem#1\endcsname}%
\let\auto@bib@innerbib\@empty
\bibitem [{\citenamefont {Essig}\ \emph {et~al.}(2013)\citenamefont {Essig} \emph {et~al.}}]{Essig:2013lka}%
  \BibitemOpen
  \bibfield  {author} {\bibinfo {author} {\bibfnamefont {R.}~\bibnamefont {Essig}} \emph {et~al.},\ }in\ \href@noop {} {\emph {\bibinfo {booktitle} {{Snowmass 2013}: {Snowmass on the Mississippi}}}}\ (\bibinfo {year} {2013})\ \Eprint {http://arxiv.org/abs/1311.0029} {arXiv:1311.0029 [hep-ph]} \BibitemShut {NoStop}%
\bibitem [{\citenamefont {Beacham}\ \emph {et~al.}(2020)\citenamefont {Beacham} \emph {et~al.}}]{Beacham:2019nyx}%
  \BibitemOpen
  \bibfield  {author} {\bibinfo {author} {\bibfnamefont {J.}~\bibnamefont {Beacham}} \emph {et~al.},\ }\href {\doibase 10.1088/1361-6471/ab4cd2} {\bibfield  {journal} {\bibinfo  {journal} {J. Phys. G}\ }\textbf {\bibinfo {volume} {47}},\ \bibinfo {pages} {010501} (\bibinfo {year} {2020})},\ \Eprint {http://arxiv.org/abs/1901.09966} {arXiv:1901.09966 [hep-ex]} \BibitemShut {NoStop}%
\bibitem [{\citenamefont {Agrawal}\ \emph {et~al.}(2021)\citenamefont {Agrawal} \emph {et~al.}}]{Agrawal:2021dbo}%
  \BibitemOpen
  \bibfield  {author} {\bibinfo {author} {\bibfnamefont {P.}~\bibnamefont {Agrawal}} \emph {et~al.},\ }\href {\doibase 10.1140/epjc/s10052-021-09703-7} {\bibfield  {journal} {\bibinfo  {journal} {Eur. Phys. J. C}\ }\textbf {\bibinfo {volume} {81}},\ \bibinfo {pages} {1015} (\bibinfo {year} {2021})},\ \Eprint {http://arxiv.org/abs/2102.12143} {arXiv:2102.12143 [hep-ph]} \BibitemShut {NoStop}%
\bibitem [{\citenamefont {Antel}\ \emph {et~al.}(2023)\citenamefont {Antel} \emph {et~al.}}]{Antel:2023hkf}%
  \BibitemOpen
  \bibfield  {author} {\bibinfo {author} {\bibfnamefont {C.}~\bibnamefont {Antel}} \emph {et~al.},\ }\href {\doibase 10.1140/epjc/s10052-023-12168-5} {\bibfield  {journal} {\bibinfo  {journal} {Eur. Phys. J. C}\ }\textbf {\bibinfo {volume} {83}},\ \bibinfo {pages} {1122} (\bibinfo {year} {2023})},\ \Eprint {http://arxiv.org/abs/2305.01715} {arXiv:2305.01715 [hep-ph]} \BibitemShut {NoStop}%
\bibitem [{\citenamefont {Chen}\ \emph {et~al.}(2024)\citenamefont {Chen}, \citenamefont {Du}, \citenamefont {Sun}, \citenamefont {Wang}, \citenamefont {Xie}, \citenamefont {Xie}, \citenamefont {Yang}, \citenamefont {Yu},\ and\ \citenamefont {Zhang}}]{Chen:2024fzk}%
  \BibitemOpen
  \bibfield  {author} {\bibinfo {author} {\bibfnamefont {L.}~\bibnamefont {Chen}}, \bibinfo {author} {\bibfnamefont {M.}~\bibnamefont {Du}}, \bibinfo {author} {\bibfnamefont {Z.}~\bibnamefont {Sun}}, \bibinfo {author} {\bibfnamefont {Z.~S.}\ \bibnamefont {Wang}}, \bibinfo {author} {\bibfnamefont {F.}~\bibnamefont {Xie}}, \bibinfo {author} {\bibfnamefont {J.-J.}\ \bibnamefont {Xie}}, \bibinfo {author} {\bibfnamefont {L.}~\bibnamefont {Yang}}, \bibinfo {author} {\bibfnamefont {P.}~\bibnamefont {Yu}}, \ and\ \bibinfo {author} {\bibfnamefont {Y.}~\bibnamefont {Zhang}},\ }\href@noop {} {\  (\bibinfo {year} {2024})},\ \Eprint {http://arxiv.org/abs/2412.09132} {arXiv:2412.09132 [hep-ph]} \BibitemShut {NoStop}%
\bibitem [{\citenamefont {Agostinelli}\ \emph {et~al.}(2003)\citenamefont {Agostinelli} \emph {et~al.}}]{GEANT4:2002zbu}%
  \BibitemOpen
  \bibfield  {author} {\bibinfo {author} {\bibfnamefont {S.}~\bibnamefont {Agostinelli}} \emph {et~al.} (\bibinfo {collaboration} {GEANT4}),\ }\href {\doibase 10.1016/S0168-9002(03)01368-8} {\bibfield  {journal} {\bibinfo  {journal} {Nucl. Instrum. Meth. A}\ }\textbf {\bibinfo {volume} {506}},\ \bibinfo {pages} {250} (\bibinfo {year} {2003})}\BibitemShut {NoStop}%
\bibitem [{\citenamefont {Wright}\ and\ \citenamefont {Kelsey}(2015)}]{Wright:2015xia}%
  \BibitemOpen
  \bibfield  {author} {\bibinfo {author} {\bibfnamefont {D.~H.}\ \bibnamefont {Wright}}\ and\ \bibinfo {author} {\bibfnamefont {M.~H.}\ \bibnamefont {Kelsey}},\ }\href {\doibase 10.1016/j.nima.2015.09.058} {\bibfield  {journal} {\bibinfo  {journal} {Nucl. Instrum. Meth. A}\ }\textbf {\bibinfo {volume} {804}},\ \bibinfo {pages} {175} (\bibinfo {year} {2015})}\BibitemShut {NoStop}%
\bibitem [{\citenamefont {Akimov}\ \emph {et~al.}(2022)\citenamefont {Akimov} \emph {et~al.}}]{PhysRevD.106.032003}%
  \BibitemOpen
  \bibfield  {author} {\bibinfo {author} {\bibfnamefont {D.}~\bibnamefont {Akimov}} \emph {et~al.} (\bibinfo {collaboration} {COHERENT Collaboration}),\ }\href {\doibase 10.1103/PhysRevD.106.032003} {\bibfield  {journal} {\bibinfo  {journal} {Phys. Rev. D}\ }\textbf {\bibinfo {volume} {106}},\ \bibinfo {pages} {032003} (\bibinfo {year} {2022})}\BibitemShut {NoStop}%
\bibitem [{\citenamefont {Folger}\ \emph {et~al.}(2004)\citenamefont {Folger}, \citenamefont {Ivanchenko},\ and\ \citenamefont {Wellisch}}]{Folger:2004zma}%
  \BibitemOpen
  \bibfield  {author} {\bibinfo {author} {\bibfnamefont {G.}~\bibnamefont {Folger}}, \bibinfo {author} {\bibfnamefont {V.~N.}\ \bibnamefont {Ivanchenko}}, \ and\ \bibinfo {author} {\bibfnamefont {J.~P.}\ \bibnamefont {Wellisch}},\ }\href {\doibase 10.1140/epja/i2003-10219-7} {\bibfield  {journal} {\bibinfo  {journal} {Eur. Phys. J. A}\ }\textbf {\bibinfo {volume} {21}},\ \bibinfo {pages} {407} (\bibinfo {year} {2004})}\BibitemShut {NoStop}%
\bibitem [{\citenamefont {Ivantchenko}\ \emph {et~al.}(2011)\citenamefont {Ivantchenko}, \citenamefont {Ivanchenko}, \citenamefont {Quesada~Molina},\ and\ \citenamefont {Incerti}}]{Ivantchenko:2011qbbc}%
  \BibitemOpen
  \bibfield  {author} {\bibinfo {author} {\bibfnamefont {A.}~\bibnamefont {Ivantchenko}}, \bibinfo {author} {\bibfnamefont {V.}~\bibnamefont {Ivanchenko}}, \bibinfo {author} {\bibfnamefont {J.}~\bibnamefont {Quesada~Molina}}, \ and\ \bibinfo {author} {\bibfnamefont {S.}~\bibnamefont {Incerti}},\ }\href {\doibase 10.3109/09553002.2011.610865} {\bibfield  {journal} {\bibinfo  {journal} {Int. J. Radiat. Biol.}\ }\textbf {\bibinfo {volume} {88}},\ \bibinfo {pages} {171} (\bibinfo {year} {2011})}\BibitemShut {NoStop}%
\bibitem [{\citenamefont {Boudard}\ \emph {et~al.}(2013)\citenamefont {Boudard}, \citenamefont {Cugnon}, \citenamefont {David}, \citenamefont {Leray},\ and\ \citenamefont {Mancusi}}]{Boudard:2012wc}%
  \BibitemOpen
  \bibfield  {author} {\bibinfo {author} {\bibfnamefont {A.}~\bibnamefont {Boudard}}, \bibinfo {author} {\bibfnamefont {J.}~\bibnamefont {Cugnon}}, \bibinfo {author} {\bibfnamefont {J.-C.}\ \bibnamefont {David}}, \bibinfo {author} {\bibfnamefont {S.}~\bibnamefont {Leray}}, \ and\ \bibinfo {author} {\bibfnamefont {D.}~\bibnamefont {Mancusi}},\ }\href {\doibase 10.1103/PhysRevC.87.014606} {\bibfield  {journal} {\bibinfo  {journal} {Phys. Rev. C}\ }\textbf {\bibinfo {volume} {87}},\ \bibinfo {pages} {014606} (\bibinfo {year} {2013})}\BibitemShut {NoStop}%
\bibitem [{\citenamefont {Bethe}(1930)}]{Bethe:1930ku}%
  \BibitemOpen
  \bibfield  {author} {\bibinfo {author} {\bibfnamefont {H.}~\bibnamefont {Bethe}},\ }\href {\doibase 10.1002/andp.19303970303} {\bibfield  {journal} {\bibinfo  {journal} {Annalen Phys.}\ }\textbf {\bibinfo {volume} {5}},\ \bibinfo {pages} {325} (\bibinfo {year} {1930})}\BibitemShut {NoStop}%
\bibitem [{\citenamefont {Livingston}\ and\ \citenamefont {Bethe}(1937)}]{RevModPhys.9.245}%
  \BibitemOpen
  \bibfield  {author} {\bibinfo {author} {\bibfnamefont {M.~S.}\ \bibnamefont {Livingston}}\ and\ \bibinfo {author} {\bibfnamefont {H.~A.}\ \bibnamefont {Bethe}},\ }\href {\doibase 10.1103/RevModPhys.9.245} {\bibfield  {journal} {\bibinfo  {journal} {Rev. Mod. Phys.}\ }\textbf {\bibinfo {volume} {9}},\ \bibinfo {pages} {245} (\bibinfo {year} {1937})}\BibitemShut {NoStop}%
\bibitem [{\citenamefont {Navas}\ \emph {et~al.}(2024)\citenamefont {Navas} \emph {et~al.}}]{ParticleDataGroup:2024cfk}%
  \BibitemOpen
  \bibfield  {author} {\bibinfo {author} {\bibfnamefont {S.}~\bibnamefont {Navas}} \emph {et~al.} (\bibinfo {collaboration} {Particle Data Group}),\ }\href {\doibase 10.1103/PhysRevD.110.030001} {\bibfield  {journal} {\bibinfo  {journal} {Phys. Rev. D}\ }\textbf {\bibinfo {volume} {110}},\ \bibinfo {pages} {030001} (\bibinfo {year} {2024})}\BibitemShut {NoStop}%
\bibitem [{\citenamefont {Sternheimer}\ \emph {et~al.}(1984)\citenamefont {Sternheimer}, \citenamefont {Berger},\ and\ \citenamefont {Seltzer}}]{Sternheimer:1983mb}%
  \BibitemOpen
  \bibfield  {author} {\bibinfo {author} {\bibfnamefont {R.~M.}\ \bibnamefont {Sternheimer}}, \bibinfo {author} {\bibfnamefont {M.~J.}\ \bibnamefont {Berger}}, \ and\ \bibinfo {author} {\bibfnamefont {S.~M.}\ \bibnamefont {Seltzer}},\ }\href {\doibase 10.1016/0092-640X(84)90002-0} {\bibfield  {journal} {\bibinfo  {journal} {Atom. Data Nucl. Data Tabl.}\ }\textbf {\bibinfo {volume} {30}},\ \bibinfo {pages} {261} (\bibinfo {year} {1984})}\BibitemShut {NoStop}%
\bibitem [{\citenamefont {Atoian}\ \emph {et~al.}(2008)\citenamefont {Atoian} \emph {et~al.}}]{Atoian:2007up}%
  \BibitemOpen
  \bibfield  {author} {\bibinfo {author} {\bibfnamefont {G.~S.}\ \bibnamefont {Atoian}} \emph {et~al.},\ }\href {\doibase 10.1016/j.nima.2007.10.022} {\bibfield  {journal} {\bibinfo  {journal} {Nucl. Instrum. Meth. A}\ }\textbf {\bibinfo {volume} {584}},\ \bibinfo {pages} {291} (\bibinfo {year} {2008})},\ \Eprint {http://arxiv.org/abs/0709.4514} {arXiv:0709.4514 [physics.ins-det]} \BibitemShut {NoStop}%
\bibitem [{\citenamefont {Ball}\ \emph {et~al.}(2020)\citenamefont {Ball} \emph {et~al.}}]{Ball:2020dnx}%
  \BibitemOpen
  \bibfield  {author} {\bibinfo {author} {\bibfnamefont {A.}~\bibnamefont {Ball}} \emph {et~al.},\ }\href {\doibase 10.1103/PhysRevD.102.032002} {\bibfield  {journal} {\bibinfo  {journal} {Phys. Rev. D}\ }\textbf {\bibinfo {volume} {102}},\ \bibinfo {pages} {032002} (\bibinfo {year} {2020})},\ \Eprint {http://arxiv.org/abs/2005.06518} {arXiv:2005.06518 [hep-ex]} \BibitemShut {NoStop}%
\bibitem [{\citenamefont {Harnik}\ \emph {et~al.}(2019)\citenamefont {Harnik}, \citenamefont {Liu},\ and\ \citenamefont {Palamara}}]{Harnik:2019zee}%
  \BibitemOpen
  \bibfield  {author} {\bibinfo {author} {\bibfnamefont {R.}~\bibnamefont {Harnik}}, \bibinfo {author} {\bibfnamefont {Z.}~\bibnamefont {Liu}}, \ and\ \bibinfo {author} {\bibfnamefont {O.}~\bibnamefont {Palamara}},\ }\href {\doibase 10.1007/JHEP07(2019)170} {\bibfield  {journal} {\bibinfo  {journal} {JHEP}\ }\textbf {\bibinfo {volume} {07}},\ \bibinfo {pages} {170} (\bibinfo {year} {2019})},\ \Eprint {http://arxiv.org/abs/1902.03246} {arXiv:1902.03246 [hep-ph]} \BibitemShut {NoStop}%
\bibitem [{\citenamefont {Blondel}\ \emph {et~al.}(2018)\citenamefont {Blondel} \emph {et~al.}}]{Blondel:2017orl}%
  \BibitemOpen
  \bibfield  {author} {\bibinfo {author} {\bibfnamefont {A.}~\bibnamefont {Blondel}} \emph {et~al.},\ }\href {\doibase 10.1088/1748-0221/13/02/P02006} {\bibfield  {journal} {\bibinfo  {journal} {JINST}\ }\textbf {\bibinfo {volume} {13}},\ \bibinfo {pages} {P02006} (\bibinfo {year} {2018})},\ \Eprint {http://arxiv.org/abs/1707.01785} {arXiv:1707.01785 [physics.ins-det]} \BibitemShut {NoStop}%
\bibitem [{\citenamefont {Kudenko}(2025)}]{Kudenko:2025dlg}%
  \BibitemOpen
  \bibfield  {author} {\bibinfo {author} {\bibfnamefont {Y.}~\bibnamefont {Kudenko}} (\bibinfo {collaboration} {T2K}),\ }\href {\doibase 10.54546/naturalscirev.100304} {\bibfield  {journal} {\bibinfo  {journal} {Natural Sci. Rev.}\ }\textbf {\bibinfo {volume} {3}},\ \bibinfo {pages} {100304} (\bibinfo {year} {2025})}\BibitemShut {NoStop}%
\bibitem [{\citenamefont {Morris}\ \emph {et~al.}(1976)\citenamefont {Morris}, \citenamefont {Mahaney},\ and\ \citenamefont {Huber}}]{doi:10.1021/j100550a010}%
  \BibitemOpen
  \bibfield  {author} {\bibinfo {author} {\bibfnamefont {J.~V.}\ \bibnamefont {Morris}}, \bibinfo {author} {\bibfnamefont {M.~A.}\ \bibnamefont {Mahaney}}, \ and\ \bibinfo {author} {\bibfnamefont {J.~R.}\ \bibnamefont {Huber}},\ }\href {\doibase 10.1021/j100550a010} {\bibfield  {journal} {\bibinfo  {journal} {The Journal of Physical Chemistry}\ }\textbf {\bibinfo {volume} {80}},\ \bibinfo {pages} {969} (\bibinfo {year} {1976})}\BibitemShut {NoStop}%
\bibitem [{\citenamefont {Biller}\ \emph {et~al.}(2020)\citenamefont {Biller}, \citenamefont {Leming},\ and\ \citenamefont {Paton}}]{Biller:2020uoi}%
  \BibitemOpen
  \bibfield  {author} {\bibinfo {author} {\bibfnamefont {S.~D.}\ \bibnamefont {Biller}}, \bibinfo {author} {\bibfnamefont {E.~J.}\ \bibnamefont {Leming}}, \ and\ \bibinfo {author} {\bibfnamefont {J.~L.}\ \bibnamefont {Paton}},\ }\href {\doibase 10.1016/j.nima.2020.164106} {\bibfield  {journal} {\bibinfo  {journal} {Nucl. Instrum. Meth. A}\ }\textbf {\bibinfo {volume} {972}},\ \bibinfo {pages} {164106} (\bibinfo {year} {2020})},\ \Eprint {http://arxiv.org/abs/2001.10825} {arXiv:2001.10825 [physics.ins-det]} \BibitemShut {NoStop}%
\bibitem [{\citenamefont {Bauer}\ \emph {et~al.}(2017)\citenamefont {Bauer}, \citenamefont {Neubert},\ and\ \citenamefont {Thamm}}]{Bauer:2017ris}%
  \BibitemOpen
  \bibfield  {author} {\bibinfo {author} {\bibfnamefont {M.}~\bibnamefont {Bauer}}, \bibinfo {author} {\bibfnamefont {M.}~\bibnamefont {Neubert}}, \ and\ \bibinfo {author} {\bibfnamefont {A.}~\bibnamefont {Thamm}},\ }\href {\doibase 10.1007/JHEP12(2017)044} {\bibfield  {journal} {\bibinfo  {journal} {JHEP}\ }\textbf {\bibinfo {volume} {12}},\ \bibinfo {pages} {044} (\bibinfo {year} {2017})},\ \Eprint {http://arxiv.org/abs/1708.00443} {arXiv:1708.00443 [hep-ph]} \BibitemShut {NoStop}%
\bibitem [{\citenamefont {Irastorza}\ and\ \citenamefont {Redondo}(2018)}]{Irastorza:2018dyq}%
  \BibitemOpen
  \bibfield  {author} {\bibinfo {author} {\bibfnamefont {I.~G.}\ \bibnamefont {Irastorza}}\ and\ \bibinfo {author} {\bibfnamefont {J.}~\bibnamefont {Redondo}},\ }\href {\doibase 10.1016/j.ppnp.2018.05.003} {\bibfield  {journal} {\bibinfo  {journal} {Prog. Part. Nucl. Phys.}\ }\textbf {\bibinfo {volume} {102}},\ \bibinfo {pages} {89} (\bibinfo {year} {2018})},\ \Eprint {http://arxiv.org/abs/1801.08127} {arXiv:1801.08127 [hep-ph]} \BibitemShut {NoStop}%
\bibitem [{\citenamefont {Brignole}\ \emph {et~al.}(1999)\citenamefont {Brignole}, \citenamefont {Perazzi},\ and\ \citenamefont {Zwirner}}]{Brignole:1999gf}%
  \BibitemOpen
  \bibfield  {author} {\bibinfo {author} {\bibfnamefont {A.}~\bibnamefont {Brignole}}, \bibinfo {author} {\bibfnamefont {E.}~\bibnamefont {Perazzi}}, \ and\ \bibinfo {author} {\bibfnamefont {F.}~\bibnamefont {Zwirner}},\ }\href {\doibase 10.1088/1126-6708/1999/09/002} {\bibfield  {journal} {\bibinfo  {journal} {JHEP}\ }\textbf {\bibinfo {volume} {09}},\ \bibinfo {pages} {002} (\bibinfo {year} {1999})},\ \Eprint {http://arxiv.org/abs/hep-ph/9904367} {arXiv:hep-ph/9904367} \BibitemShut {NoStop}%
\bibitem [{\citenamefont {Gorbunov}(2001)}]{Gorbunov:2000th}%
  \BibitemOpen
  \bibfield  {author} {\bibinfo {author} {\bibfnamefont {D.~S.}\ \bibnamefont {Gorbunov}},\ }\href {\doibase 10.1016/S0550-3213(01)00122-5} {\bibfield  {journal} {\bibinfo  {journal} {Nucl. Phys. B}\ }\textbf {\bibinfo {volume} {602}},\ \bibinfo {pages} {213} (\bibinfo {year} {2001})},\ \Eprint {http://arxiv.org/abs/hep-ph/0007325} {arXiv:hep-ph/0007325} \BibitemShut {NoStop}%
\bibitem [{\citenamefont {Georgi}\ \emph {et~al.}(1986)\citenamefont {Georgi}, \citenamefont {Kaplan},\ and\ \citenamefont {Randall}}]{Georgi:1986df}%
  \BibitemOpen
  \bibfield  {author} {\bibinfo {author} {\bibfnamefont {H.}~\bibnamefont {Georgi}}, \bibinfo {author} {\bibfnamefont {D.~B.}\ \bibnamefont {Kaplan}}, \ and\ \bibinfo {author} {\bibfnamefont {L.}~\bibnamefont {Randall}},\ }\href {\doibase 10.1016/0370-2693(86)90688-X} {\bibfield  {journal} {\bibinfo  {journal} {Phys. Lett. B}\ }\textbf {\bibinfo {volume} {169}},\ \bibinfo {pages} {73} (\bibinfo {year} {1986})}\BibitemShut {NoStop}%
\bibitem [{\citenamefont {Aloni}\ \emph {et~al.}(2019)\citenamefont {Aloni}, \citenamefont {Soreq},\ and\ \citenamefont {Williams}}]{Aloni:2018vki}%
  \BibitemOpen
  \bibfield  {author} {\bibinfo {author} {\bibfnamefont {D.}~\bibnamefont {Aloni}}, \bibinfo {author} {\bibfnamefont {Y.}~\bibnamefont {Soreq}}, \ and\ \bibinfo {author} {\bibfnamefont {M.}~\bibnamefont {Williams}},\ }\href {\doibase 10.1103/PhysRevLett.123.031803} {\bibfield  {journal} {\bibinfo  {journal} {Phys. Rev. Lett.}\ }\textbf {\bibinfo {volume} {123}},\ \bibinfo {pages} {031803} (\bibinfo {year} {2019})},\ \Eprint {http://arxiv.org/abs/1811.03474} {arXiv:1811.03474 [hep-ph]} \BibitemShut {NoStop}%
\bibitem [{\citenamefont {Bauer}\ \emph {et~al.}(2021{\natexlab{a}})\citenamefont {Bauer}, \citenamefont {Neubert}, \citenamefont {Renner}, \citenamefont {Schnubel},\ and\ \citenamefont {Thamm}}]{Bauer:2020jbp}%
  \BibitemOpen
  \bibfield  {author} {\bibinfo {author} {\bibfnamefont {M.}~\bibnamefont {Bauer}}, \bibinfo {author} {\bibfnamefont {M.}~\bibnamefont {Neubert}}, \bibinfo {author} {\bibfnamefont {S.}~\bibnamefont {Renner}}, \bibinfo {author} {\bibfnamefont {M.}~\bibnamefont {Schnubel}}, \ and\ \bibinfo {author} {\bibfnamefont {A.}~\bibnamefont {Thamm}},\ }\href {\doibase 10.1007/JHEP04(2021)063} {\bibfield  {journal} {\bibinfo  {journal} {JHEP}\ }\textbf {\bibinfo {volume} {04}},\ \bibinfo {pages} {063} (\bibinfo {year} {2021}{\natexlab{a}})},\ \Eprint {http://arxiv.org/abs/2012.12272} {arXiv:2012.12272 [hep-ph]} \BibitemShut {NoStop}%
\bibitem [{\citenamefont {Bauer}\ \emph {et~al.}(2021{\natexlab{b}})\citenamefont {Bauer}, \citenamefont {Neubert}, \citenamefont {Renner}, \citenamefont {Schnubel},\ and\ \citenamefont {Thamm}}]{Bauer:2021wjo}%
  \BibitemOpen
  \bibfield  {author} {\bibinfo {author} {\bibfnamefont {M.}~\bibnamefont {Bauer}}, \bibinfo {author} {\bibfnamefont {M.}~\bibnamefont {Neubert}}, \bibinfo {author} {\bibfnamefont {S.}~\bibnamefont {Renner}}, \bibinfo {author} {\bibfnamefont {M.}~\bibnamefont {Schnubel}}, \ and\ \bibinfo {author} {\bibfnamefont {A.}~\bibnamefont {Thamm}},\ }\href {\doibase 10.1103/PhysRevLett.127.081803} {\bibfield  {journal} {\bibinfo  {journal} {Phys. Rev. Lett.}\ }\textbf {\bibinfo {volume} {127}},\ \bibinfo {pages} {081803} (\bibinfo {year} {2021}{\natexlab{b}})},\ \Eprint {http://arxiv.org/abs/2102.13112} {arXiv:2102.13112 [hep-ph]} \BibitemShut {NoStop}%
\bibitem [{\citenamefont {Jerhot}\ \emph {et~al.}(2022)\citenamefont {Jerhot}, \citenamefont {D\"obrich}, \citenamefont {Ertas}, \citenamefont {Kahlhoefer},\ and\ \citenamefont {Spadaro}}]{Jerhot:2022chi}%
  \BibitemOpen
  \bibfield  {author} {\bibinfo {author} {\bibfnamefont {J.}~\bibnamefont {Jerhot}}, \bibinfo {author} {\bibfnamefont {B.}~\bibnamefont {D\"obrich}}, \bibinfo {author} {\bibfnamefont {F.}~\bibnamefont {Ertas}}, \bibinfo {author} {\bibfnamefont {F.}~\bibnamefont {Kahlhoefer}}, \ and\ \bibinfo {author} {\bibfnamefont {T.}~\bibnamefont {Spadaro}},\ }\href {\doibase 10.1007/JHEP07(2022)094} {\bibfield  {journal} {\bibinfo  {journal} {JHEP}\ }\textbf {\bibinfo {volume} {07}},\ \bibinfo {pages} {094} (\bibinfo {year} {2022})},\ \Eprint {http://arxiv.org/abs/2201.05170} {arXiv:2201.05170 [hep-ph]} \BibitemShut {NoStop}%
\bibitem [{\citenamefont {Dalla Valle~Garcia}\ \emph {et~al.}(2024)\citenamefont {Dalla Valle~Garcia}, \citenamefont {Kahlhoefer}, \citenamefont {Ovchynnikov},\ and\ \citenamefont {Zaporozhchenko}}]{DallaValleGarcia:2023xhh}%
  \BibitemOpen
  \bibfield  {author} {\bibinfo {author} {\bibfnamefont {G.}~\bibnamefont {Dalla Valle~Garcia}}, \bibinfo {author} {\bibfnamefont {F.}~\bibnamefont {Kahlhoefer}}, \bibinfo {author} {\bibfnamefont {M.}~\bibnamefont {Ovchynnikov}}, \ and\ \bibinfo {author} {\bibfnamefont {A.}~\bibnamefont {Zaporozhchenko}},\ }\href {\doibase 10.1103/PhysRevD.109.055042} {\bibfield  {journal} {\bibinfo  {journal} {Phys. Rev. D}\ }\textbf {\bibinfo {volume} {109}},\ \bibinfo {pages} {055042} (\bibinfo {year} {2024})},\ \Eprint {http://arxiv.org/abs/2310.03524} {arXiv:2310.03524 [hep-ph]} \BibitemShut {NoStop}%
\bibitem [{\citenamefont {Ovchynnikov}\ and\ \citenamefont {Zaporozhchenko}(2025)}]{Ovchynnikov:2025gpx}%
  \BibitemOpen
  \bibfield  {author} {\bibinfo {author} {\bibfnamefont {M.}~\bibnamefont {Ovchynnikov}}\ and\ \bibinfo {author} {\bibfnamefont {A.}~\bibnamefont {Zaporozhchenko}},\ }\href@noop {} {\  (\bibinfo {year} {2025})},\ \Eprint {http://arxiv.org/abs/2501.04525} {arXiv:2501.04525 [hep-ph]} \BibitemShut {NoStop}%
\bibitem [{\citenamefont {Cortina~Gil}\ \emph {et~al.}(2021)\citenamefont {Cortina~Gil} \emph {et~al.}}]{NA62:2020xlg}%
  \BibitemOpen
  \bibfield  {author} {\bibinfo {author} {\bibfnamefont {E.}~\bibnamefont {Cortina~Gil}} \emph {et~al.} (\bibinfo {collaboration} {NA62}),\ }\href {\doibase 10.1007/JHEP03(2021)058} {\bibfield  {journal} {\bibinfo  {journal} {JHEP}\ }\textbf {\bibinfo {volume} {03}},\ \bibinfo {pages} {058} (\bibinfo {year} {2021})},\ \Eprint {http://arxiv.org/abs/2011.11329} {arXiv:2011.11329 [hep-ex]} \BibitemShut {NoStop}%
\bibitem [{\citenamefont {Artamonov}\ \emph {et~al.}(2009)\citenamefont {Artamonov} \emph {et~al.}}]{BNL-E949:2009dza}%
  \BibitemOpen
  \bibfield  {author} {\bibinfo {author} {\bibfnamefont {A.~V.}\ \bibnamefont {Artamonov}} \emph {et~al.} (\bibinfo {collaboration} {BNL-E949}),\ }\href {\doibase 10.1103/PhysRevD.79.092004} {\bibfield  {journal} {\bibinfo  {journal} {Phys. Rev. D}\ }\textbf {\bibinfo {volume} {79}},\ \bibinfo {pages} {092004} (\bibinfo {year} {2009})},\ \Eprint {http://arxiv.org/abs/0903.0030} {arXiv:0903.0030 [hep-ex]} \BibitemShut {NoStop}%
\bibitem [{\citenamefont {Bergsma}\ \emph {et~al.}(1985{\natexlab{a}})\citenamefont {Bergsma} \emph {et~al.}}]{CHARM:1985anb}%
  \BibitemOpen
  \bibfield  {author} {\bibinfo {author} {\bibfnamefont {F.}~\bibnamefont {Bergsma}} \emph {et~al.} (\bibinfo {collaboration} {CHARM}),\ }\href {\doibase 10.1016/0370-2693(85)90400-9} {\bibfield  {journal} {\bibinfo  {journal} {Phys. Lett. B}\ }\textbf {\bibinfo {volume} {157}},\ \bibinfo {pages} {458} (\bibinfo {year} {1985}{\natexlab{a}})}\BibitemShut {NoStop}%
\bibitem [{\citenamefont {Bl{\"u}mlein}\ \emph {et~al.}(1991)\citenamefont {Bl{\"u}mlein} \emph {et~al.}}]{Blumlein:1990ay}%
  \BibitemOpen
  \bibfield  {author} {\bibinfo {author} {\bibfnamefont {J.}~\bibnamefont {Bl{\"u}mlein}} \emph {et~al.},\ }\href {\doibase 10.1007/BF01548556} {\bibfield  {journal} {\bibinfo  {journal} {Z. Phys.}\ }\textbf {\bibinfo {volume} {C51}},\ \bibinfo {pages} {341} (\bibinfo {year} {1991})}\BibitemShut {NoStop}%
\bibitem [{\citenamefont {Blumlein}\ \emph {et~al.}(1992)\citenamefont {Blumlein} \emph {et~al.}}]{Blumlein:1991xh}%
  \BibitemOpen
  \bibfield  {author} {\bibinfo {author} {\bibfnamefont {J.}~\bibnamefont {Blumlein}} \emph {et~al.},\ }\href {\doibase 10.1142/S0217751X9200171X} {\bibfield  {journal} {\bibinfo  {journal} {Int. J. Mod. Phys. A}\ }\textbf {\bibinfo {volume} {7}},\ \bibinfo {pages} {3835} (\bibinfo {year} {1992})}\BibitemShut {NoStop}%
\bibitem [{\citenamefont {Tulin}(2014)}]{Tulin:2014tya}%
  \BibitemOpen
  \bibfield  {author} {\bibinfo {author} {\bibfnamefont {S.}~\bibnamefont {Tulin}},\ }\href {\doibase 10.1103/PhysRevD.89.114008} {\bibfield  {journal} {\bibinfo  {journal} {Phys. Rev. D}\ }\textbf {\bibinfo {volume} {89}},\ \bibinfo {pages} {114008} (\bibinfo {year} {2014})},\ \Eprint {http://arxiv.org/abs/1404.4370} {arXiv:1404.4370 [hep-ph]} \BibitemShut {NoStop}%
\bibitem [{\citenamefont {Curtin}\ \emph {et~al.}(2023)\citenamefont {Curtin}, \citenamefont {Kahn},\ and\ \citenamefont {Nguyen}}]{Curtin:2023bcf}%
  \BibitemOpen
  \bibfield  {author} {\bibinfo {author} {\bibfnamefont {D.}~\bibnamefont {Curtin}}, \bibinfo {author} {\bibfnamefont {Y.}~\bibnamefont {Kahn}}, \ and\ \bibinfo {author} {\bibfnamefont {R.}~\bibnamefont {Nguyen}},\ }\href {\doibase 10.1103/PhysRevD.108.095039} {\bibfield  {journal} {\bibinfo  {journal} {Phys. Rev. D}\ }\textbf {\bibinfo {volume} {108}},\ \bibinfo {pages} {095039} (\bibinfo {year} {2023})},\ \Eprint {http://arxiv.org/abs/2305.19309} {arXiv:2305.19309 [hep-ph]} \BibitemShut {NoStop}%
\bibitem [{\citenamefont {Hemmert}\ \emph {et~al.}(1998)\citenamefont {Hemmert}, \citenamefont {Holstein},\ and\ \citenamefont {Kambor}}]{Hemmert:1997ye}%
  \BibitemOpen
  \bibfield  {author} {\bibinfo {author} {\bibfnamefont {T.~R.}\ \bibnamefont {Hemmert}}, \bibinfo {author} {\bibfnamefont {B.~R.}\ \bibnamefont {Holstein}}, \ and\ \bibinfo {author} {\bibfnamefont {J.}~\bibnamefont {Kambor}},\ }\href {\doibase 10.1088/0954-3899/24/10/003} {\bibfield  {journal} {\bibinfo  {journal} {J. Phys. G}\ }\textbf {\bibinfo {volume} {24}},\ \bibinfo {pages} {1831} (\bibinfo {year} {1998})},\ \Eprint {http://arxiv.org/abs/hep-ph/9712496} {arXiv:hep-ph/9712496} \BibitemShut {NoStop}%
\bibitem [{\citenamefont {Fettes}\ \emph {et~al.}(2000)\citenamefont {Fettes}, \citenamefont {Meissner}, \citenamefont {Mojzis},\ and\ \citenamefont {Steininger}}]{Fettes:2000gb}%
  \BibitemOpen
  \bibfield  {author} {\bibinfo {author} {\bibfnamefont {N.}~\bibnamefont {Fettes}}, \bibinfo {author} {\bibfnamefont {U.-G.}\ \bibnamefont {Meissner}}, \bibinfo {author} {\bibfnamefont {M.}~\bibnamefont {Mojzis}}, \ and\ \bibinfo {author} {\bibfnamefont {S.}~\bibnamefont {Steininger}},\ }\href {\doibase 10.1006/aphy.2000.6059} {\bibfield  {journal} {\bibinfo  {journal} {Annals Phys.}\ }\textbf {\bibinfo {volume} {283}},\ \bibinfo {pages} {273} (\bibinfo {year} {2000})},\ \bibinfo {note} {[Erratum: Annals Phys. 288, 249--250 (2001)]},\ \Eprint {http://arxiv.org/abs/hep-ph/0001308} {arXiv:hep-ph/0001308} \BibitemShut {NoStop}%
\bibitem [{\citenamefont {Rijneveen}\ \emph {et~al.}(2022)\citenamefont {Rijneveen}, \citenamefont {Gasparyan}, \citenamefont {Krebs},\ and\ \citenamefont {Epelbaum}}]{Rijneveen:2021bfw}%
  \BibitemOpen
  \bibfield  {author} {\bibinfo {author} {\bibfnamefont {N.}~\bibnamefont {Rijneveen}}, \bibinfo {author} {\bibfnamefont {A.~M.}\ \bibnamefont {Gasparyan}}, \bibinfo {author} {\bibfnamefont {H.}~\bibnamefont {Krebs}}, \ and\ \bibinfo {author} {\bibfnamefont {E.}~\bibnamefont {Epelbaum}},\ }\href {\doibase 10.1103/PhysRevC.106.025202} {\bibfield  {journal} {\bibinfo  {journal} {Phys. Rev. C}\ }\textbf {\bibinfo {volume} {106}},\ \bibinfo {pages} {025202} (\bibinfo {year} {2022})},\ \Eprint {http://arxiv.org/abs/2108.01619} {arXiv:2108.01619 [nucl-th]} \BibitemShut {NoStop}%
\bibitem [{\citenamefont {Ellis}(2017)}]{Ellis:2016jkw}%
  \BibitemOpen
  \bibfield  {author} {\bibinfo {author} {\bibfnamefont {J.}~\bibnamefont {Ellis}},\ }\href {\doibase 10.1016/j.cpc.2016.08.019} {\bibfield  {journal} {\bibinfo  {journal} {Comput. Phys. Commun.}\ }\textbf {\bibinfo {volume} {210}},\ \bibinfo {pages} {103} (\bibinfo {year} {2017})},\ \Eprint {http://arxiv.org/abs/1601.05437} {arXiv:1601.05437 [hep-ph]} \BibitemShut {NoStop}%
\bibitem [{\citenamefont {Shin}\ and\ \citenamefont {Yun}(2023)}]{Shin:2022ulh}%
  \BibitemOpen
  \bibfield  {author} {\bibinfo {author} {\bibfnamefont {C.~S.}\ \bibnamefont {Shin}}\ and\ \bibinfo {author} {\bibfnamefont {S.}~\bibnamefont {Yun}},\ }\href {\doibase 10.1103/PhysRevD.108.055014} {\bibfield  {journal} {\bibinfo  {journal} {Phys. Rev. D}\ }\textbf {\bibinfo {volume} {108}},\ \bibinfo {pages} {055014} (\bibinfo {year} {2023})},\ \Eprint {http://arxiv.org/abs/2211.15677} {arXiv:2211.15677 [hep-ph]} \BibitemShut {NoStop}%
\bibitem [{\citenamefont {Mertig}\ \emph {et~al.}(1991)\citenamefont {Mertig}, \citenamefont {Bohm},\ and\ \citenamefont {Denner}}]{Mertig:1990an}%
  \BibitemOpen
  \bibfield  {author} {\bibinfo {author} {\bibfnamefont {R.}~\bibnamefont {Mertig}}, \bibinfo {author} {\bibfnamefont {M.}~\bibnamefont {Bohm}}, \ and\ \bibinfo {author} {\bibfnamefont {A.}~\bibnamefont {Denner}},\ }\href {\doibase 10.1016/0010-4655(91)90130-D} {\bibfield  {journal} {\bibinfo  {journal} {Comput. Phys. Commun.}\ }\textbf {\bibinfo {volume} {64}},\ \bibinfo {pages} {345} (\bibinfo {year} {1991})}\BibitemShut {NoStop}%
\bibitem [{\citenamefont {Shtabovenko}\ \emph {et~al.}(2016)\citenamefont {Shtabovenko}, \citenamefont {Mertig},\ and\ \citenamefont {Orellana}}]{Shtabovenko:2016sxi}%
  \BibitemOpen
  \bibfield  {author} {\bibinfo {author} {\bibfnamefont {V.}~\bibnamefont {Shtabovenko}}, \bibinfo {author} {\bibfnamefont {R.}~\bibnamefont {Mertig}}, \ and\ \bibinfo {author} {\bibfnamefont {F.}~\bibnamefont {Orellana}},\ }\href {\doibase 10.1016/j.cpc.2016.06.008} {\bibfield  {journal} {\bibinfo  {journal} {Comput. Phys. Commun.}\ }\textbf {\bibinfo {volume} {207}},\ \bibinfo {pages} {432} (\bibinfo {year} {2016})},\ \Eprint {http://arxiv.org/abs/1601.01167} {arXiv:1601.01167 [hep-ph]} \BibitemShut {NoStop}%
\bibitem [{\citenamefont {Shtabovenko}\ \emph {et~al.}(2020)\citenamefont {Shtabovenko}, \citenamefont {Mertig},\ and\ \citenamefont {Orellana}}]{Shtabovenko:2020gxv}%
  \BibitemOpen
  \bibfield  {author} {\bibinfo {author} {\bibfnamefont {V.}~\bibnamefont {Shtabovenko}}, \bibinfo {author} {\bibfnamefont {R.}~\bibnamefont {Mertig}}, \ and\ \bibinfo {author} {\bibfnamefont {F.}~\bibnamefont {Orellana}},\ }\href {\doibase 10.1016/j.cpc.2020.107478} {\bibfield  {journal} {\bibinfo  {journal} {Comput. Phys. Commun.}\ }\textbf {\bibinfo {volume} {256}},\ \bibinfo {pages} {107478} (\bibinfo {year} {2020})},\ \Eprint {http://arxiv.org/abs/2001.04407} {arXiv:2001.04407 [hep-ph]} \BibitemShut {NoStop}%
\bibitem [{\citenamefont {Shin}\ and\ \citenamefont {Yun}(2022)}]{Shin:2021bvz}%
  \BibitemOpen
  \bibfield  {author} {\bibinfo {author} {\bibfnamefont {C.~S.}\ \bibnamefont {Shin}}\ and\ \bibinfo {author} {\bibfnamefont {S.}~\bibnamefont {Yun}},\ }\href {\doibase 10.1007/JHEP02(2022)133} {\bibfield  {journal} {\bibinfo  {journal} {JHEP}\ }\textbf {\bibinfo {volume} {02}},\ \bibinfo {pages} {133} (\bibinfo {year} {2022})},\ \Eprint {http://arxiv.org/abs/2110.03362} {arXiv:2110.03362 [hep-ph]} \BibitemShut {NoStop}%
\bibitem [{\citenamefont {Pukhov}\ \emph {et~al.}(1999)\citenamefont {Pukhov}, \citenamefont {Boos}, \citenamefont {Dubinin}, \citenamefont {Edneral}, \citenamefont {Ilyin}, \citenamefont {Kovalenko}, \citenamefont {Kryukov}, \citenamefont {Savrin}, \citenamefont {Shichanin},\ and\ \citenamefont {Semenov}}]{Pukhov:1999gg}%
  \BibitemOpen
  \bibfield  {author} {\bibinfo {author} {\bibfnamefont {A.}~\bibnamefont {Pukhov}}, \bibinfo {author} {\bibfnamefont {E.}~\bibnamefont {Boos}}, \bibinfo {author} {\bibfnamefont {M.}~\bibnamefont {Dubinin}}, \bibinfo {author} {\bibfnamefont {V.}~\bibnamefont {Edneral}}, \bibinfo {author} {\bibfnamefont {V.}~\bibnamefont {Ilyin}}, \bibinfo {author} {\bibfnamefont {D.}~\bibnamefont {Kovalenko}}, \bibinfo {author} {\bibfnamefont {A.}~\bibnamefont {Kryukov}}, \bibinfo {author} {\bibfnamefont {V.}~\bibnamefont {Savrin}}, \bibinfo {author} {\bibfnamefont {S.}~\bibnamefont {Shichanin}}, \ and\ \bibinfo {author} {\bibfnamefont {A.}~\bibnamefont {Semenov}},\ }\href@noop {} {\  (\bibinfo {year} {1999})},\ \Eprint {http://arxiv.org/abs/hep-ph/9908288} {arXiv:hep-ph/9908288} \BibitemShut {NoStop}%
\bibitem [{\citenamefont {Boos}\ \emph {et~al.}(2004)\citenamefont {Boos}, \citenamefont {Bunichev}, \citenamefont {Dubinin}, \citenamefont {Dudko}, \citenamefont {Ilyin}, \citenamefont {Kryukov}, \citenamefont {Edneral}, \citenamefont {Savrin}, \citenamefont {Semenov},\ and\ \citenamefont {Sherstnev}}]{CompHEP:2004qpa}%
  \BibitemOpen
  \bibfield  {author} {\bibinfo {author} {\bibfnamefont {E.}~\bibnamefont {Boos}}, \bibinfo {author} {\bibfnamefont {V.}~\bibnamefont {Bunichev}}, \bibinfo {author} {\bibfnamefont {M.}~\bibnamefont {Dubinin}}, \bibinfo {author} {\bibfnamefont {L.}~\bibnamefont {Dudko}}, \bibinfo {author} {\bibfnamefont {V.}~\bibnamefont {Ilyin}}, \bibinfo {author} {\bibfnamefont {A.}~\bibnamefont {Kryukov}}, \bibinfo {author} {\bibfnamefont {V.}~\bibnamefont {Edneral}}, \bibinfo {author} {\bibfnamefont {V.}~\bibnamefont {Savrin}}, \bibinfo {author} {\bibfnamefont {A.}~\bibnamefont {Semenov}}, \ and\ \bibinfo {author} {\bibfnamefont {A.}~\bibnamefont {Sherstnev}} (\bibinfo {collaboration} {CompHEP}),\ }\href {\doibase 10.1016/j.nima.2004.07.096} {\bibfield  {journal} {\bibinfo  {journal} {Nucl. Instrum. Meth. A}\ }\textbf {\bibinfo {volume} {534}},\ \bibinfo {pages} {250} (\bibinfo {year} {2004})},\ \Eprint {http://arxiv.org/abs/hep-ph/0403113} {arXiv:hep-ph/0403113} \BibitemShut {NoStop}%
\bibitem [{\citenamefont {Ilten}\ \emph {et~al.}(2018)\citenamefont {Ilten}, \citenamefont {Soreq}, \citenamefont {Williams},\ and\ \citenamefont {Xue}}]{Ilten:2018crw}%
  \BibitemOpen
  \bibfield  {author} {\bibinfo {author} {\bibfnamefont {P.}~\bibnamefont {Ilten}}, \bibinfo {author} {\bibfnamefont {Y.}~\bibnamefont {Soreq}}, \bibinfo {author} {\bibfnamefont {M.}~\bibnamefont {Williams}}, \ and\ \bibinfo {author} {\bibfnamefont {W.}~\bibnamefont {Xue}},\ }\href {\doibase 10.1007/JHEP06(2018)004} {\bibfield  {journal} {\bibinfo  {journal} {JHEP}\ }\textbf {\bibinfo {volume} {06}},\ \bibinfo {pages} {004} (\bibinfo {year} {2018})},\ \Eprint {http://arxiv.org/abs/1801.04847} {arXiv:1801.04847 [hep-ph]} \BibitemShut {NoStop}%
\bibitem [{\citenamefont {Cortina~Gil}\ \emph {et~al.}(2023)\citenamefont {Cortina~Gil} \emph {et~al.}}]{NA62:2023qyn}%
  \BibitemOpen
  \bibfield  {author} {\bibinfo {author} {\bibfnamefont {E.}~\bibnamefont {Cortina~Gil}} \emph {et~al.} (\bibinfo {collaboration} {NA62}),\ }\href {\doibase 10.1007/JHEP09(2023)035} {\bibfield  {journal} {\bibinfo  {journal} {JHEP}\ }\textbf {\bibinfo {volume} {09}},\ \bibinfo {pages} {035} (\bibinfo {year} {2023})},\ \Eprint {http://arxiv.org/abs/2303.08666} {arXiv:2303.08666 [hep-ex]} \BibitemShut {NoStop}%
\bibitem [{\citenamefont {Astier}\ \emph {et~al.}(2001)\citenamefont {Astier} \emph {et~al.}}]{Astier:2001ck}%
  \BibitemOpen
  \bibfield  {author} {\bibinfo {author} {\bibfnamefont {P.}~\bibnamefont {Astier}} \emph {et~al.} (\bibinfo {collaboration} {NOMAD}),\ }\href {\doibase 10.1016/S0370-2693(01)00362-8} {\bibfield  {journal} {\bibinfo  {journal} {Phys. Lett.}\ }\textbf {\bibinfo {volume} {B506}},\ \bibinfo {pages} {27} (\bibinfo {year} {2001})},\ \Eprint {http://arxiv.org/abs/hep-ex/0101041} {arXiv:hep-ex/0101041 [hep-ex]} \BibitemShut {NoStop}%
\bibitem [{\citenamefont {Tsai}\ \emph {et~al.}(2021)\citenamefont {Tsai}, \citenamefont {deNiverville},\ and\ \citenamefont {Liu}}]{Tsai:2019buq}%
  \BibitemOpen
  \bibfield  {author} {\bibinfo {author} {\bibfnamefont {Y.-D.}\ \bibnamefont {Tsai}}, \bibinfo {author} {\bibfnamefont {P.}~\bibnamefont {deNiverville}}, \ and\ \bibinfo {author} {\bibfnamefont {M.~X.}\ \bibnamefont {Liu}},\ }\href {\doibase 10.1103/PhysRevLett.126.181801} {\bibfield  {journal} {\bibinfo  {journal} {Phys. Rev. Lett.}\ }\textbf {\bibinfo {volume} {126}},\ \bibinfo {pages} {181801} (\bibinfo {year} {2021})},\ \Eprint {http://arxiv.org/abs/1908.07525} {arXiv:1908.07525 [hep-ph]} \BibitemShut {NoStop}%
\bibitem [{\citenamefont {Bernardi}\ \emph {et~al.}(1986)\citenamefont {Bernardi} \emph {et~al.}}]{Bernardi:1985ny}%
  \BibitemOpen
  \bibfield  {author} {\bibinfo {author} {\bibfnamefont {G.}~\bibnamefont {Bernardi}} \emph {et~al.},\ }\href {\doibase 10.1016/0370-2693(86)91602-3} {\bibfield  {journal} {\bibinfo  {journal} {Phys. Lett.}\ }\textbf {\bibinfo {volume} {B166}},\ \bibinfo {pages} {479} (\bibinfo {year} {1986})}\BibitemShut {NoStop}%
\bibitem [{\citenamefont {Bergsma}\ \emph {et~al.}(1985{\natexlab{b}})\citenamefont {Bergsma} \emph {et~al.}}]{Bergsma:1985qz}%
  \BibitemOpen
  \bibfield  {author} {\bibinfo {author} {\bibfnamefont {F.}~\bibnamefont {Bergsma}} \emph {et~al.} (\bibinfo {collaboration} {CHARM}),\ }\href {\doibase 10.1016/0370-2693(85)90400-9} {\bibfield  {journal} {\bibinfo  {journal} {Phys. Lett.}\ }\textbf {\bibinfo {volume} {157B}},\ \bibinfo {pages} {458} (\bibinfo {year} {1985}{\natexlab{b}})}\BibitemShut {NoStop}%
\bibitem [{\citenamefont {Foguel}\ \emph {et~al.}(2022)\citenamefont {Foguel}, \citenamefont {Reimitz},\ and\ \citenamefont {Funchal}}]{Foguel:2022ppx}%
  \BibitemOpen
  \bibfield  {author} {\bibinfo {author} {\bibfnamefont {A.~L.}\ \bibnamefont {Foguel}}, \bibinfo {author} {\bibfnamefont {P.}~\bibnamefont {Reimitz}}, \ and\ \bibinfo {author} {\bibfnamefont {R.~Z.}\ \bibnamefont {Funchal}},\ }\href {\doibase 10.1007/JHEP04(2022)119} {\bibfield  {journal} {\bibinfo  {journal} {JHEP}\ }\textbf {\bibinfo {volume} {04}},\ \bibinfo {pages} {119} (\bibinfo {year} {2022})},\ \Eprint {http://arxiv.org/abs/2201.01788} {arXiv:2201.01788 [hep-ph]} \BibitemShut {NoStop}%
\bibitem [{\citenamefont {Batell}\ \emph {et~al.}(2009)\citenamefont {Batell}, \citenamefont {Pospelov},\ and\ \citenamefont {Ritz}}]{Batell:2009di}%
  \BibitemOpen
  \bibfield  {author} {\bibinfo {author} {\bibfnamefont {B.}~\bibnamefont {Batell}}, \bibinfo {author} {\bibfnamefont {M.}~\bibnamefont {Pospelov}}, \ and\ \bibinfo {author} {\bibfnamefont {A.}~\bibnamefont {Ritz}},\ }\href {\doibase 10.1103/PhysRevD.80.095024} {\bibfield  {journal} {\bibinfo  {journal} {Phys. Rev. D}\ }\textbf {\bibinfo {volume} {80}},\ \bibinfo {pages} {095024} (\bibinfo {year} {2009})},\ \Eprint {http://arxiv.org/abs/0906.5614} {arXiv:0906.5614 [hep-ph]} \BibitemShut {NoStop}%
\bibitem [{\citenamefont {Foot}(1991)}]{Foot:1990mn}%
  \BibitemOpen
  \bibfield  {author} {\bibinfo {author} {\bibfnamefont {R.}~\bibnamefont {Foot}},\ }\href {\doibase 10.1142/S0217732391000543} {\bibfield  {journal} {\bibinfo  {journal} {Mod. Phys. Lett. A}\ }\textbf {\bibinfo {volume} {6}},\ \bibinfo {pages} {527} (\bibinfo {year} {1991})}\BibitemShut {NoStop}%
\bibitem [{\citenamefont {Holdom}(1986)}]{Holdom:1985ag}%
  \BibitemOpen
  \bibfield  {author} {\bibinfo {author} {\bibfnamefont {B.}~\bibnamefont {Holdom}},\ }\href {\doibase 10.1016/0370-2693(86)91377-8} {\bibfield  {journal} {\bibinfo  {journal} {Phys. Lett. B}\ }\textbf {\bibinfo {volume} {166}},\ \bibinfo {pages} {196} (\bibinfo {year} {1986})}\BibitemShut {NoStop}%
\bibitem [{\citenamefont {Pich}(1998)}]{pich1998effective}%
  \BibitemOpen
  \bibfield  {author} {\bibinfo {author} {\bibfnamefont {A.}~\bibnamefont {Pich}},\ }\href@noop {} {\enquote {\bibinfo {title} {Effective field theory},}\ } (\bibinfo {year} {1998}),\ \Eprint {http://arxiv.org/abs/hep-ph/9806303} {arXiv:hep-ph/9806303 [hep-ph]} \BibitemShut {NoStop}%
\bibitem [{\citenamefont {Acciarri}\ \emph {et~al.}(2020)\citenamefont {Acciarri} \emph {et~al.}}]{ArgoNeuT:2019ckq}%
  \BibitemOpen
  \bibfield  {author} {\bibinfo {author} {\bibfnamefont {R.}~\bibnamefont {Acciarri}} \emph {et~al.} (\bibinfo {collaboration} {ArgoNeuT}),\ }\href {\doibase 10.1103/PhysRevLett.124.131801} {\bibfield  {journal} {\bibinfo  {journal} {Phys. Rev. Lett.}\ }\textbf {\bibinfo {volume} {124}},\ \bibinfo {pages} {131801} (\bibinfo {year} {2020})},\ \Eprint {http://arxiv.org/abs/1911.07996} {arXiv:1911.07996 [hep-ex]} \BibitemShut {NoStop}%
\bibitem [{\citenamefont {Gorbunov}\ \emph {et~al.}(2021)\citenamefont {Gorbunov}, \citenamefont {Krasnov}, \citenamefont {Kudenko},\ and\ \citenamefont {Suvorov}}]{Gorbunov:2021jog}%
  \BibitemOpen
  \bibfield  {author} {\bibinfo {author} {\bibfnamefont {D.}~\bibnamefont {Gorbunov}}, \bibinfo {author} {\bibfnamefont {I.}~\bibnamefont {Krasnov}}, \bibinfo {author} {\bibfnamefont {Y.}~\bibnamefont {Kudenko}}, \ and\ \bibinfo {author} {\bibfnamefont {S.}~\bibnamefont {Suvorov}},\ }\href {\doibase 10.1016/j.physletb.2021.136641} {\bibfield  {journal} {\bibinfo  {journal} {Phys. Lett. B}\ }\textbf {\bibinfo {volume} {822}},\ \bibinfo {pages} {136641} (\bibinfo {year} {2021})},\ \Eprint {http://arxiv.org/abs/2103.11814} {arXiv:2103.11814 [hep-ph]} \BibitemShut {NoStop}%
\bibitem [{\citenamefont {Zyla}\ \emph {et~al.}(2020)\citenamefont {Zyla} \emph {et~al.}}]{ParticleDataGroup:2020ssz}%
  \BibitemOpen
  \bibfield  {author} {\bibinfo {author} {\bibfnamefont {P.~A.}\ \bibnamefont {Zyla}} \emph {et~al.} (\bibinfo {collaboration} {Particle Data Group}),\ }\href {\doibase 10.1093/ptep/ptaa104} {\bibfield  {journal} {\bibinfo  {journal} {PTEP}\ }\textbf {\bibinfo {volume} {2020}},\ \bibinfo {pages} {083C01} (\bibinfo {year} {2020})}\BibitemShut {NoStop}%
\bibitem [{\citenamefont {Uehling}(1954)}]{Uehling:1954wp}%
  \BibitemOpen
  \bibfield  {author} {\bibinfo {author} {\bibfnamefont {E.~A.}\ \bibnamefont {Uehling}},\ }\href {\doibase 10.1146/annurev.ns.04.120154.001531} {\bibfield  {journal} {\bibinfo  {journal} {Ann. Rev. Nucl. Part. Sci.}\ }\textbf {\bibinfo {volume} {4}},\ \bibinfo {pages} {315} (\bibinfo {year} {1954})}\BibitemShut {NoStop}%
\bibitem [{\citenamefont {Arefyeva}\ \emph {et~al.}(2022)\citenamefont {Arefyeva}, \citenamefont {Gninenko}, \citenamefont {Gorbunov},\ and\ \citenamefont {Kirpichnikov}}]{Arefyeva:2022eba}%
  \BibitemOpen
  \bibfield  {author} {\bibinfo {author} {\bibfnamefont {N.}~\bibnamefont {Arefyeva}}, \bibinfo {author} {\bibfnamefont {S.}~\bibnamefont {Gninenko}}, \bibinfo {author} {\bibfnamefont {D.}~\bibnamefont {Gorbunov}}, \ and\ \bibinfo {author} {\bibfnamefont {D.}~\bibnamefont {Kirpichnikov}},\ }\href {\doibase 10.1103/PhysRevD.106.035029} {\bibfield  {journal} {\bibinfo  {journal} {Phys. Rev. D}\ }\textbf {\bibinfo {volume} {106}},\ \bibinfo {pages} {035029} (\bibinfo {year} {2022})},\ \Eprint {http://arxiv.org/abs/2204.03984} {arXiv:2204.03984 [hep-ph]} \BibitemShut {NoStop}%
\bibitem [{\citenamefont {Magill}\ \emph {et~al.}(2019)\citenamefont {Magill}, \citenamefont {Plestid}, \citenamefont {Pospelov},\ and\ \citenamefont {Tsai}}]{PhysRevLett.122.071801}%
  \BibitemOpen
  \bibfield  {author} {\bibinfo {author} {\bibfnamefont {G.}~\bibnamefont {Magill}}, \bibinfo {author} {\bibfnamefont {R.}~\bibnamefont {Plestid}}, \bibinfo {author} {\bibfnamefont {M.}~\bibnamefont {Pospelov}}, \ and\ \bibinfo {author} {\bibfnamefont {Y.-D.}\ \bibnamefont {Tsai}},\ }\href {\doibase 10.1103/PhysRevLett.122.071801} {\bibfield  {journal} {\bibinfo  {journal} {Phys. Rev. Lett.}\ }\textbf {\bibinfo {volume} {122}},\ \bibinfo {pages} {071801} (\bibinfo {year} {2019})}\BibitemShut {NoStop}%
\bibitem [{\citenamefont {Marocco}\ and\ \citenamefont {Sarkar}(2021)}]{10.21468/SciPostPhys.10.2.043}%
  \BibitemOpen
  \bibfield  {author} {\bibinfo {author} {\bibfnamefont {G.}~\bibnamefont {Marocco}}\ and\ \bibinfo {author} {\bibfnamefont {S.}~\bibnamefont {Sarkar}},\ }\href {\doibase 10.21468/SciPostPhys.10.2.043} {\bibfield  {journal} {\bibinfo  {journal} {SciPost Phys.}\ }\textbf {\bibinfo {volume} {10}},\ \bibinfo {pages} {043} (\bibinfo {year} {2021})}\BibitemShut {NoStop}%
\bibitem [{\citenamefont {Barak}\ \emph {et~al.}(2024)\citenamefont {Barak} \emph {et~al.}}]{SENSEI:2023gie}%
  \BibitemOpen
  \bibfield  {author} {\bibinfo {author} {\bibfnamefont {L.}~\bibnamefont {Barak}} \emph {et~al.} (\bibinfo {collaboration} {SENSEI}),\ }\href {\doibase 10.1103/PhysRevLett.133.071801} {\bibfield  {journal} {\bibinfo  {journal} {Phys. Rev. Lett.}\ }\textbf {\bibinfo {volume} {133}},\ \bibinfo {pages} {071801} (\bibinfo {year} {2024})},\ \Eprint {http://arxiv.org/abs/2305.04964} {arXiv:2305.04964 [hep-ex]} \BibitemShut {NoStop}%
\end{thebibliography}%

\end{document}